\documentclass[12pt]{article}

\usepackage[export]{adjustbox}

\usepackage{scicite}
\usepackage{graphicx}
\usepackage{graphicx,booktabs,tabularx,graphics}
\usepackage{tabulary}
\usepackage{adjustbox}

\usepackage{times}
\usepackage{color}
\usepackage[table,x11names,dvipsnames]{xcolor}
\usepackage{colortbl}
\usepackage{hyperref}

\usepackage{times}
\usepackage{subfigure}
\usepackage{amsmath}

\newenvironment{sciabstract}{%
\begin{quote} \bf}
{\end{quote}}

\newcounter{lastnote}

\title{Predicting success in the worldwide start-up network}

\author
{Moreno Bonaventura$^{1,2,\dagger}$, Valerio Ciotti$^{1,2,\dagger}$, Pietro Panzarasa$^{2}$\\  Silvia Liverani$^{1,3}$,  Lucas Lacasa$^1$, Vito Latora$^{1,3,4,5}$\\
\normalsize{$^{1}$School of Mathematical Sciences, Queen Mary University of London,}\\
\normalsize{Mile End Road, E14NS, London (UK)}\\
\normalsize{$^{2}$School of Business and Management, Queen Mary University of London,}\\
\normalsize{Mile End Road, E14NS, London (UK).}\\
\normalsize{$^{3}$The Alan Turing Institute, The British Library NW12DB, London  (UK)}\\
\normalsize{$^{4}$Dipartimento di Fisica e Astronomia, Universit\`a di Catania and INFN, 95123 Catania (Italy)}\\
\normalsize{$^{5}$Complexity Science Hub Vienna (CSHV), Vienna (Austria)}\\
\normalsize{$^{\dagger}$ M.B. and V.C. contributed equally to this work.}
}

\date{}

\begin{document} 

\baselineskip24pt
\maketitle

\begin{sciabstract}
  By drawing on large-scale online data we construct
    and analyze the time-varying worldwide network of professional
    relationships among start-ups. The nodes of this network represent
    companies, while the links model the flow of employees and the
    associated transfer of know-how across companies. We use network
    centrality measures to assess, at an early stage, the likelihood
    of the long-term positive performance of a start-up,
    showing that the start-up
    network has predictive power and provides valuable recommendations
    doubling the current state of the art performance of venture
    funds. Our network-based approach not only offers an effective
    alternative to the labour-intensive screening processes of venture
    capital firms, but can also enable entrepreneurs and policy-makers
    to conduct a more objective assessment of the long-term potentials  
    of innovation ecosystems and to target interventions accordingly.
\end{sciabstract}


\noindent Recent years have witnessed an unprecedented growth of
interest in start-up companies. Policy-makers have been keen to
sustain young entrepreneurs' innovative efforts with a view to
injecting new driving forces into the economy and foster job creation
and technological advancements \cite{eu2012, USeconomicReport2016,
  Haltiwanger2013, Mazzucato2013}. Investors have been lured by the
opportunity of disproportionally high returns typically associated
with radical new developments and technological discontinuities. Large
corporations have relied on various forms of external collaborations
with newly established firms to outsource innovation processes and
stay abreast of technological breakthroughs
\cite{Chesbrough2003}. Undoubtedly, knowledge-intensive ventures such
as start-ups can have a large positive impact on the economy and
society. Yet they typically suffer from a liability of newness
\cite{Freeman}, and cannot avoid the uncertainties and sunk costs
resulting from disruptive product developments, uncharted markets and
rapidly changing technological regimes \cite{Powell2005}. For these
reasons, their long-term benefits are inherently difficult to predict,
and their economic net present value cannot be unambiguously assessed
\cite{Shane2008}.
 
\noindent Indeed traditional models of business evaluation, based on
historical trends of data (e.g., on sales, production capacity,
internal growth, and markets size) are mostly inapplicable to
start-ups, chiefly because their limited history does not provide
sufficient data. Venture capitalists and private investors often
evaluate start-ups primarily based on the qualifications and dexterity
of the entrepreneurs, on their potential to create new markets or
niches and to unleash the ``gales of creative destruction''
\cite{Schumpeter}. The process of screening and evaluating companies
in their early stages is therefore a subjective and labor-intensive
task, and is inevitably fraught with biases and uncertainty.

 \noindent To overcoming these limitation, we
   propose a novel and data-driven framework for assessing the
   long-term economic potential of newly established start-ups. Our
   study draws upon the construction and analysis of the worldwide
   network of professional relationships among start-ups. Such network
   provides the backbone and the channels through which knowledge can be
   gained, transferred, shared, and recombined. For instance, skilled
   employees moving across firms in search of novel opportunities 
   can bring with them know-how on cutting-edge
   technologies, advisors who gained experience in one firm can help
   identify the most effective strategies in another, whilst well
   connected investors, lenders and board members can rely on the
   knowledge gained in one firm to tap business and funding
   opportunities in another.

\noindent Previous work has investigated how knowledge transfer
impacts upon the performance of start-ups; yet information flows have
been simply inferred mainly through data on patents \cite{Guzman2015},
interorganizational collaborations \cite{Powell1996}, co-location of
firms and their proximity to universities \cite{Saxenian1996}. Other
studies have analyzed social networks (e.g., inventor collaboration
networks, interlocking directorates) to unveil the microscopic level
of interactions among individuals; yet their scope has been limited
mostly to specific industries or small geographic areas, and to a
fairly small observation period \cite{Powell1996,Sorenson2006,
  Ferrary2009}. Owing to lack of data, what still remains to be
studied is the global network underpinning knowledge exchange in the
worldwide innovation ecosystem. Equally, the competitive advantage of
differential information-rich network positions and their role in
opening up, expediting, or obstructing pathways to firms' long-term
success have been left largely unexplored.

\vspace{5mm}


\begin{figure}
	\centering
		\includegraphics[width=1.1\textwidth,center]{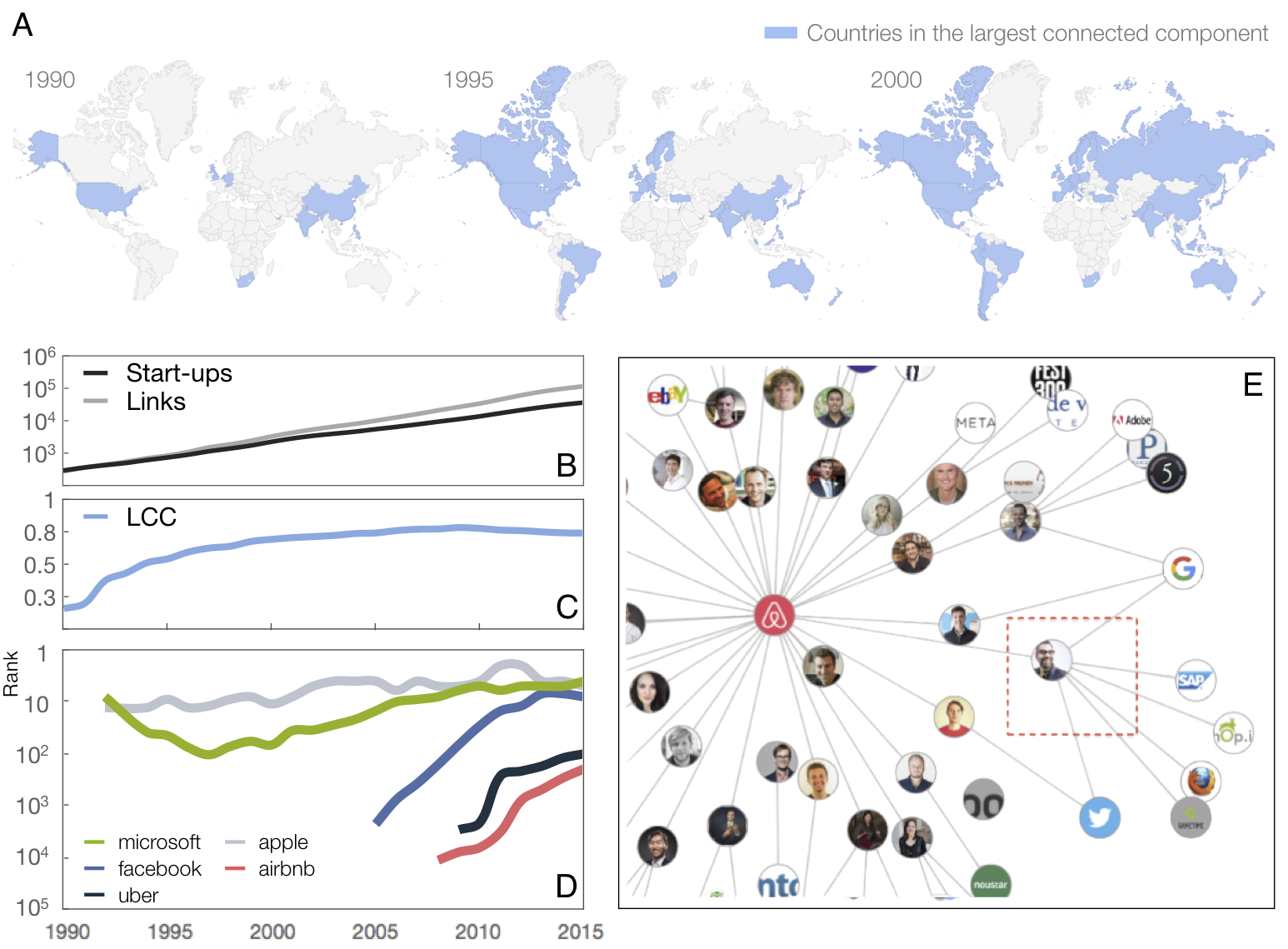}
		\caption{{\bf The time-varying network of professional relationships among start-ups.} (A) Countries that, over time, joined the largest connected component (LCC) of the worldwide start-up (WWS) network are highlighted in blue. (B) Evolution over time of the number of firms and links in the WWS network. (C) Evolution over time of the fraction of nodes in the LCC. (D) Evolution over time of the closeness centrality rank of five popular firms. (E) Airbnb's ego-centered network.
}

\end{figure}

\noindent \textbf{The world-wide network of start-ups.}
Here we study the complex time-varying network \cite{Latora2017,Masuda2016} of interactions among all start-ups in the worldwide
innovation ecosystem over a period of 26 years (1990-2015). To this end,
we collected all data on firms and people (i.e., founders, employees,
advisors, investors, and board members) available from the
\textit{www.crunchbase.com} website. Drawing on the data, we
first constructed a bipartite graph in which people
are connected to start-ups according to their professional role. We
then obtained the projected one-mode time-varying graph in which
start-ups are the nodes and two companies are connected when they
share at least one individual that plays or has played a professional
role in both companies (see Supplementary Material (SM) for details).
At the micro scale, employees working in a
  company can perceive the intrinsic value of new appealing
  opportunities and switch companies accordingly. This mobility
  creates an intel flow between companies, where those receiving
  employees increase their fitness by capitalizing on the know-how the
  employee is bringing with her. Such microscopic dynamics is thus
  captured and modelled by the creation of new edges at the level of the
  network of start-ups. As a consequence, companies which are perceived
  at the micro scale as
  appealing opportunities by mobile employees will likely boost their
  connectivity and therefore will acquire a more central position in
  the overall time-varying network.
\\
The resulting time-varying 
{\em World Wide Start-up} ({\em WWS}) network comprises
$41,830$ companies distributed across $117$ countries around the
globe, and $135,099$ links among them (see Fig S3 and S4 in the
SM). Fig 1A highlights the countries in which start-ups have joined,
over time, the largest connected component of the network
\cite{Latora2017,Masuda2016}. Fig 1B indicates that the number of
nodes and links in the WWS network has grown exponentially over the
last 26 years.
In the same period, various communities of start-ups
around the globe have joined together to form the largest connected
component including about $80\%$ of the nodes of the network (Fig
1C). Currently, an average of 4,74 ``degrees of separation'' between
any two companies characterizes the WWS network.

\noindent At the micro scale, Fig 1E shows a snapshot of the network
of interactions between Airbnb and other companies based on shared
individuals. As an illustration, in 2013 Airbnb hired Mr Thomas Arend
(highlighted in the red square), who had previously acted as a senior
product manager in Google, as an international product leader in
Twitter, and as a product manager in Mozilla. As
  previously pointed out, the professional network thus reveals the
potential flow of knowledge between Airbnb and the three other
companies in which Mr Arend had played a role. Moreover, as new links
were forged over time, the topological distances from Airbnb to all
other firms in the WWS network were reduced, which in turn enabled
Airbnb to gain new knowledge and tap business opportunities beyond its
immediate local neighborhood.

The mechanistic interpretation of employees'
  mobility inducing link creation discussed above and illustrated in Fig 1E 
suggests that the potential exposure to
  knowledge of a start-up in the WWS network, and its subsequent
  likelihood to excel in the future, should be well captured by its
  network centrality over time.
  To test this hypothesis we have considered different measures of
  node centrality \cite{Wasserman1994}.
  For parsimony here we focus on the results obtained by {\it closeness
  centrality} as it assesses the centrality of a node in the network from its
  average distance from all the other nodes, although similar
  results has also been found by some other centrality measures, such as
  betweenness or degree (see SM). In each month of the observation period, we
ranked companies according to their values of closeness centrality
(i.e., top nodes are firms with the highest closeness). Fig 1D is an
example of the large variety of observed trajectories as companies
moved towards higher or lower ranks, i.e., they obtained a larger or
smaller proximity to all other companies in the network. Notice that
Apple has always been in the Top 10 firms over the entire period,
while Microsoft exhibited an initial decline followed by a constant
rise towards the central region of the network. The trajectories of
once upon a time younger start-ups, such as Facebook, Airbnb, and
Uber, are instead characterized by an abrupt and swift move to the
highest positions of the ranking soon after their foundation, possibly
as a result of the boost in activity that has characterized the
venture capital industry in recent years.

\vspace{5mm}

\noindent \textbf{Early-stage prediction of high performance.}
To investigate the interplay
between the position of a given firm in the WWS network and its
long-term economic performance, from \textit{www.crunchbase.com} we
collected additional data on funding rounds, acquisitions, and initial
public offerings (IPOs). For each month $t$, we obtained the list of
$N(t)$ firms, ranked in terms of closeness, that can be classified as
``open deals'' for investors, namely: (i) they have not yet received
funding; (ii) they have not yet been acquired; and (iii) they have not
yet been listed in the stock exchange market (see Fig S5 in SM). As an
example, the company WhatsApp, which ranked 1,060$^{\rm th}$ in June
2009 in the full list, occupied the 15$^{\rm th}$ position in the
open-deals list in the same month. Notice that, by assessing a firm's
network position prior to any financial acquisition or IPO, our
analysis is not subject to possible biases arising from the effects
that the capital market might have upon the firm's expected
performance. 
Furthermore, predicting the long-term economic
  performance of firms in the open-deal list is arguably a challenging
  task, as illustrated by the fact that the average success of
   venture funds early-stage investments in similar open deals is only around
   10-15\% (see section S4.2 in SM).
   Over the range of 26 years of the dataset, a total of $5305$ different start-ups were identified as open-deals.
\\
Our recommendation method is based on the hypothesis
  that start-ups with higher values of closeness centrality at an early stage
  are more likely to show signs of positive long-term economic performance.
Accordingly, we counted the total
number $M(t)$ of firms inside the open-deal list that, within a time
window $\Delta t =7$ years starting at month $t$, succeeded in
securing at least one of the following positive outcomes: (i) they
took over one or more firms; (ii) they were acquired by one or more
firms; or (iii) they underwent an IPO. To assess the accuracy of our
recommendation method in early identifying successful companies, we checked
how many of the Top $n=20$ companies in the closeness-based ranking of
open-deals obtained a positive outcome (see Fig S6 in SM).
\\
\noindent Fig 2A reports the ``success rate'' $S$ (blue curve) of the
recommendation method, defined as $S(t)=m(t)/n$, where $m(t)$ is the
number of firms with a positive outcome included in the Top $n=20$
firms, and $S^{\rm rand}(t)=M(t)/N(t)$ (black curve) is the success
rate expected in the case of random ordering of companies, i.e. 
  the expected success of a null model of random sampling without replacement which complies with a hypergeometric distribution (see SM section 4 for details).
The $p$-value in the top panel of Fig 2A measures the probability of
obtaining, by chance, a success rate larger than $S(t)$, with low
values of $p$ (highlighted regions) indicating the time periods where
the prediction is statistically significant ($p$-value $ <0.05$). From
mid 2001 to mid 2004, the success rate of our
recommendation method (blue curve) is remarkably
larger than the one based on random expectations (black curve), and
the $p$-value is always smaller than 0.01. $S(t)$ exhibits an
exceptional peak of $50\%$ in June 2003 ($p$-value = 0.0001). From
2004 to 2007, the blue curve decreases, reaching a local minimum at a
time when a global financial crisis was triggered by the US housing
bubble. In this period (as well as during the collapse of the dot-com
bubble in 1999-2001), even though the success rate still exceeds
random expectations, the high $p$-values indicate that the predictions
are not statistically significant. Finally, after mid 2007, the
performance of the prediction increases, and it stabilizes around
$35\%$ ($p$-value = 0.01).  For completeness, Fig S7 in SM reports
results based on different lengths of the recommendation list and on
different time windows. The observed dependence of
  the performance of our network-based recommendation method on the level
  of external financial market stress should be studied in more depth.

\noindent In Fig 2B, we characterize the overall performance of the
recommendation method over the entire period of observation. Results
indicate that about $30\%$ of the firms appearing in the Top 20 in any
month from 2000 to 2009 have indeed achieved a positive economic outcome within 7 years since the time of our recommendation. The black error
bars indicate the expected success rates and standard deviations in
the case of random ordering of companies ($p$-values in this case are
all below $10^{-5}$). Interestingly, the 
  random null model provides an expected success rate which is indeed
  comparable to the actual performance that venture funds focusing on early-stage start-ups reach through costly and labour-intensive screening processes (see
  section 4.2 in SM for details), while the performance of our recommendation method
  is considerably superior.
\\
We further checked the robustness of our methodology by replicating
the analysis based on the Top 50 and Top 100 (reported in Fig 2B), for two additional time windows
$\Delta t = 6$ and $\Delta t = 8$ years (see Fig S8 in SM) and an
alternative method of aggregation of the success rate across the
entire observation period (see Fig S9 in SM). We
  also controlled for different confounding factors such as
  start-up size, geographical location or structural role of venture
  capital funds in the start-up network, finding that our conclusions hold (see section 5 of the SM).
\\
Finally, notice that the method presented here
  only provides a simple heuristic recommendation, i.e. it does not
  quantify the probability of {\it each} start-up
  in the open-deal list to show economic success in the future.
  In Section 6 of the SM we further
  studied this possibility by using a suite of logistic regression
  methods to {\it predict} success of each and every start-up in the open-deal
  list. We indeed found that a snapshot
  of the closeness centrality ranking of a given start-up could predict its future economic outcome  (\textsc{F1 score} $=0.6$), in qualitative
  agreement with findings in Fig 2.

\vspace{5mm}

\begin{figure}[]
\includegraphics[width=1.1\textwidth,center]{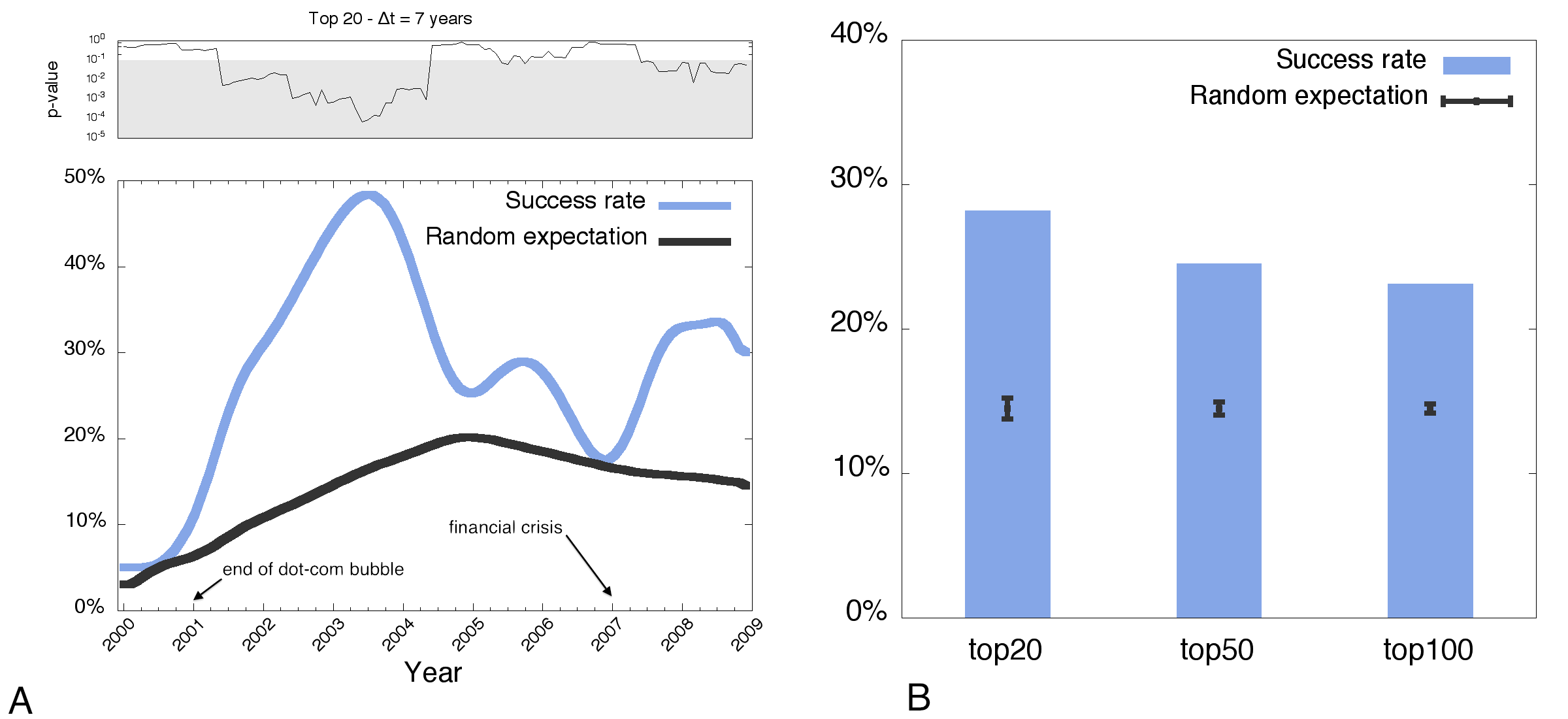}	
\caption{{\bf Closeness-based ranking of open-deals and predicting
    long-term success.} (A) The performance of our recommendation method in
  predicting companies' success on a monthly basis compared to the
   expected performance of a null model (random ordering of
  companies). The top panel reports the probability ($p$-value) of
  obtaining, by chance, a success rate larger than the one observed in
  the corresponding month.
  The gray-shaded region indicates the time periods where the
  prediction is statistically significant ($p$-value $<$ 0.05). (B)
  The overall performance of our method over the entire
  period of observation based on the Top $20$, $50$ and $100$ firms
  with the highest closeness centrality. The black error bars indicate
  the expected success rates and standard deviations in the case of
  random ordering of companies.}
\end{figure}

\noindent \textbf{Implications.} As lack of data and subjective biases
inevitably impede a proper and rigorous evaluation of risky and newly
established innovative activities, our study has indicated that the
network of professional relationships among start-ups can unlock the
long-term potential of risky ventures whose economic net present value
would otherwise be difficult to measure. Our recommendation method can
help stakeholders devise and fine-tune a number of effective
strategies, simply based on the underlying network. Employees,
business consultants, board members, bankers and lenders can identify
the opportunities with the highest long-term economic
potential. Individual and institutional investors can discern
financial deals and build appropriate portfolios that most suit their
investment preferences. Entrepreneurs can hone their networking
prowess and strategies for sustaining professional inter-firm
partnering and securing a winning streak over the long run. Finally,
governmental bodies and policy-makers can concentrate their attention
and efforts on the economic activities and geographic areas with the
most promising value-generating potential (e.g., activities with the
capacity of job creation, youth employment and skill development,
educational and technological enhancement) for both the national and
local communities. Sociological and economic research has vastly
investigated the impact of knowledge spillovers \cite{Stuart2003},
involvement in inter-firm alliances \cite{Sampson2007} and network
position \cite{Uzi1999} on firms' performance, innovation capacity,
propensity to collaborate, and growth rates. Yet, whether the
centrality in the professional network of newly established
knowledge-intensive firms can help predict their long-term economic
success has largely remained a moot question. Our work is the first
attempt to pave the way in this direction, and
  represents a contribution, from a different angle, to the ongoing
  discussion on the science of success\cite{Barabasib2018},
  complementing recent findings in different fields such as
  science\cite{Penner2013,Ma2015,Sinatra2016} and arts
  \cite{Williams2019,Fraiberger2018}.\\

{\small \noindent {\bf Acknowledgements:} 
  LL acknowledges funding from EPSRC grant EP/P01660X/1. VL
  acknowledges funding from EPSRC grant EP/N013492/1. Authors 
  express their gratitude to StartupNetwork s.r.l for providing
  data and computational infrastructure.
}

\bibliography{scibib}

\bibliographystyle{Science}

\noindent {\bf Supplementary Materials} include:\\
 Supplementary section S1) Data set: additional details\\
 Supplementary section S2) Construction of the World Wide Start-up  (WWS) network\\
Supplementary section S3) Analysis of the WWS network\\
Supplementary section S4) Open-deals recommendation method\\
Supplementary section S5) Additional analysis\\
Supplementary section S6) From recommendation to prediction of start-up success: supervised learning approaches\\
 Supplementary Tables S1--S4\\ 
 Supplementary figures Fig S3--S19\\
Supplementary References (27--37)\\

\renewcommand\theequation{{S\arabic{equation}}}
\renewcommand\thetable{{S\arabic{table}}}
\renewcommand\thefigure{{S\arabic{figure}}}
\renewcommand\thesection{{S\arabic{section}}}

\newcolumntype{a}{>{\columncolor{Gray}}c}

\definecolor{Gray}{gray}{0.85}

\renewcommand\refname{Additional References}


\newpage

\section*{\huge{\underline{\textsc{Supplementary Material}}}}
\vspace{20mm}
\section{Data set: additional details}

Data were collected from the \texttt{crunchbase.com} Web API and were
updated until December 2015. The data provided by the Crunchbase
website are manually recorded and managed by several contributors
(e.g., incubators, venture funds, individuals) affiliated with the
Crunchbase platform. Moreover, the data are further enriched by Web
crawlers that scrape the Web, on a daily basis, in search for news
about IPOs, acquisitions, and funding rounds. To date Crunchbase is
widely regarded as the world's most comprehensive open data set about
start-up companies. It contains detailed information on organizations
from all over the world and belonging to four categories, namely
companies, investors, schools, and groups. Among schools there are
$383$ universities, including top-tier institutions such as Stanford
University, the Massachusetts Institute of Technology (MIT), and many
others. In addition to people's business activity, the data track
information about their educational paths, and consequently their
access to academic knowledge.

The total number of organizations listed at the date
  of data collection amounted to $530,604$. However, a large number of
  entries contained very limited information, no profile pictures, and
  no employees' records. Accordingly, we needed to clean
  the data keeping only the organizations for which enough
  information was provided, and for which such information was
  reliable (see Section S2 for more details). This finally limited the number
  of organizations to $41,830$. For this work, all these 
  organizations were included in the construction
  of the network. Notice however, that only organizations belonging to the
  category ``companies'' and, at the same time, younger than two years,
  have been included in the recommendation list.
For each organization we extracted all the
people included in the team (e.g., founders, advisors, board member,
employees, alumni) and additional information such as details on
firms' foundation dates, locations of the firms' headquarters,
founding rounds, acquisitions, and IPOs. Organizations and people are
uniquely identified by alphanumeric IDs. All data are time-stamped,
and an accurate reconstruction of historical events was made possible
by the use of trust codes, i.e., numerical codes provided by
Crunchbase to indicate the reliability of a certain timestamp. The
timestamps indicate the dates of foundation, funding rounds,
acquisitions, and IPOs, as well as the start and the end times of job
roles.

\section{Construction of the {\em World Wide Start-up} (WWS) network}

We constructed a bipartite time-varying graph with $N_1=41,830$ nodes
representing organizations distributed across $117$ countries around
the globe, $N_2=36,278$ nodes representing people, and
$K_{12}=284,460$ links between people and organizations. The graph is
time-varying because each node and each link have an associated
timestamp, representing, respectively, the time an organization was
founded and the time a person was affiliated (and held a variety of
roles) with a given organization. Notice that in the construction of
the time-varying graph we retained only the timestamps whose trust
code guarantees the reliability of the year and month. Additionally,
we cleaned the data by solving and removing inconsistencies such as an
employee's role starting at a date prior to the company's
foundation. In these cases, we retained the most reliable information
according to the trust code value. Inconsistencies were removed by
adopting a {\it strong self-penalising data cleaning strategy}. In
particular, we did not make any assumption on dates, nor did we
attempt to infer timestamps. As a result, we do not retain in the
graph links whose timestamps cannot be determined in a reliable
way. This approach to data cleaning strengthens the validity of our
results because it ensures that companies do not gain higher positions
in the closeness centrality rank score as a result of connections that
were forged at subsequent dates to those incorrectly or only partially
reported in the data set. In this way we avoid biases that could
artificially inflate the success rate of the method, and accordingly
our results can safely be seen as conservative lower bounds.

We then projected the bipartite time-varying graph onto a one-mode
graph in which two companies are connected when they share at least
one individual that plays or has played a professional role in both
companies. Such a graph comprises $N_1=41,830$ companies and
$K=135,099$ links among them, and is here referred to as the {\em
  World Wide Start-up} ({\em WWS}) network. The projected graph is
time-varying like the original bipartite graph: a link between any two
companies is forged as soon as one individual with a professional role
in one company takes on a role in the other
company. Since the creation of
  these links denote intel transfer between companies, we
  realistically assume that such intel flow generates considerable
  know-how for the nodes receiving new links. Once created, the links are
  then maintained, since the know-how of a given
  company is not destroyed or removed.
\begin{figure}[htbp]
   \centering
   \includegraphics[width=6in]{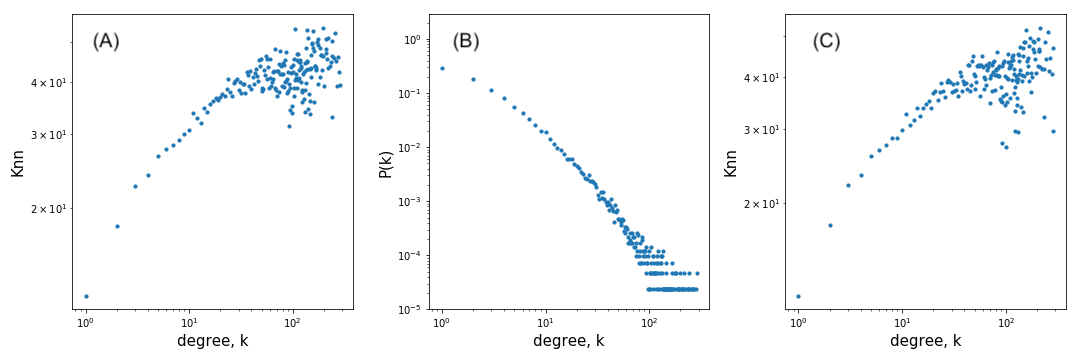} 
   \caption{{\bf Assortativity and degree distribution of the WWS
       network.}~(A) Average degree of nearest
       neighbours $k_{nn}$ for classes of nodes with degree $k$ in the
       WWS. (B) The power-law degree distribution of the WWS
       network. (C) Similar to panel (A), but where all venture
       capital firms (see Section S5.5 for a list) have been
       removed.}
   \label{fig:SI:degree_distro}
\end{figure}

\section{Analysis of the WWS network}

For completeness, we have calculated a variety of quantities for measuring the characteristics of the structure of the WWS network. In particular, from 1990 to 2015, for every month, we have computed the number of companies (nodes) and links, and examined the partition of the WWS network into distinct connected components.
A connected component of a network is a subgraph in which any two nodes are connected to each other by at least one path [{\it 15, 16}].
If the network has more than one component, one can identify the largest connected component (LCC), namely the component with the largest number of nodes.  
The countries highlighted in blue in Fig 1A (main text) are those that have at least one start-up that is part of the LCC of the WWS network.
Fig 1C (main text) shows a rapid growth in the fraction of start-ups in the LCC, thus highlighting the tendency that companies have to establish new connections with one another and move toward the core of the network.
Like many other real-world complex networks, the WWS network is characterised by  a rich topological structure, a small average shortest path length ($\ell = 4.74$), and a high value of the average clustering coefficient, $ C =0.6$, as expected from the one-mode projection of a bipartite network [{\it 15, 16}]. 
The value of the average shortest path length is similar to the one obtained for an equivalent Erd\"os-Renyi random graphs \cite{Erdos1959} with the same number of nodes and edges ($\ell ^{\rm random}$  = 4.17). However, the statistical features of the WWS differ from those characterising random graphs: the degree distribution approaches a power-law with an exponent greater than $2$ (see Fig. \ref{fig:SI:degree_distro}, panel B), the assortativity coefficient \cite{Newman2003} is positive, namely $\gamma = 0.11$ (see Fig. \ref{fig:SI:degree_distro}, panel A) and this result holds even if all venture capital firms are removed from the network (panel C). The clustering coefficient is significantly larger than the one obtained for a corresponding random network,  $C^{\rm rand} = 0.00013$.

\begin{figure}[htbp] 
   \centering
   \includegraphics[width=3in]{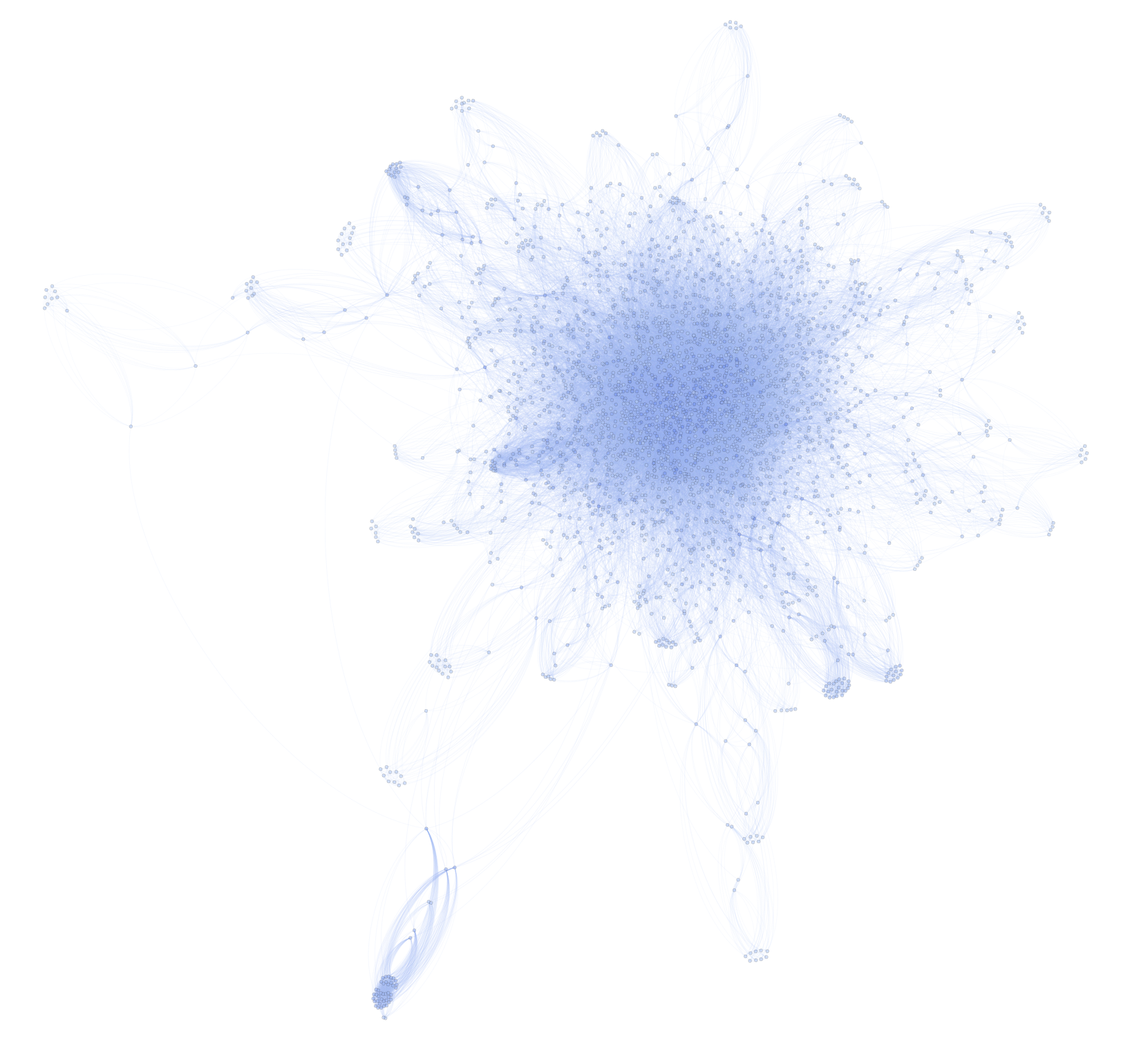} 
   \caption{\textbf{Visualisation of the WWS network}. Owing to visualisation constraints, only the 10th shell of the k-core decomposition is displayed in the image. The graph shown here includes 8\% of the nodes and 31\% of the links in the complete WWS network.}
   \label{fig:SI:graph}
\end{figure}

To offer a glimpse of the structure of the WWS network, in Fig. \ref{fig:SI:graph} we show the subgraph obtained by using the k-core decomposition technique and including only the nodes that belongs to the 10th shell. 
The k-core decomposition of a graph \cite{Erdos1966,Seidman1983,Wasserman1994} is a technique that iteratively deletes nodes starting from the most peripheral ones (i.e., nodes with degree equal to $1$) and progressively unveil the most central and interconnected core of the network. Nodes are assigned to a \textit{core value} equal to $k$ accordingly to the k-core subgraph to which they belong.

\section{Open-deals recommendation method}
\label{prediction}

Our working hypothesis is that companies with a central position in the network have higher exposure to knowledge and easier access to resources than companies with peripheral positions. If this is the case, centrally positioned companies will be better equipped to compete and have higher chances to survive, grow and flourish than peripheral ones. 
We have therefore used network centrality measures [{\it 15, 16}] that capture the structural centrality of a node in a graph, with a view to identifying companies with a large long-term economic potential. 

The concept of centrality and the related measures were first
introduced in the context of social network analysis
\cite{Wasserman1994}. 
the centrality of a company we have computed, on a monthly basis, its
closeness centrality in the WWS network. Several other
  centrality measures have also been considered, and the results are reported
  in Section S5. The closeness centrality quantifies the
importance of a node in the graph by measuring its mean distance from
all other nodes. The closeness centrality $C_{i}(t)$ of a node $i$,
$i=1,2,\ldots, {N}(t)$ is defined as:
\begin{equation}
C_{i}(t) = \frac{{N}(t) - 1} {\sum_{ j } d_{ij}(t)},\label{def_closeness}
\end{equation}
where ${N}(t)$ is the number of nodes in the graph at time $t$, while $d_{ij}(t)$ is the graph distance between the two nodes $i$ and $j$, measured as the number of links in the shortest path between the two companies. To account for disconnected components we used the generalisation of closeness centrality proposed in \cite{NanLin1976}.

Our claim is that young start-ups with
  proportionally higher values of closeness centrality will have a
  higher likelihood to become successful in later years. This
  can be readily translated into several possible heuristics
  to provide recommendation for investing into a given
  start-up. Among other possibilities, we have considered the
  following recommendation method. For each month $t$, we ranked all
the ${N}(t)$ companies according to their values of closeness
centrality $C_{i}(t)$, such that the top nodes are those with the
highest closeness. From the ranked lists we then removed the companies
that can reasonably be regarded as irrelevant deals to investors,
i.e., those companies that had already been acquired, had already been
listed in a stock market, or had received funding from other
investors. The $N(t)$ companies retained in the analysis belong to the
so-called \textit{open-deals ranked list} at month
$t$. Notice that, by definition, the open-deal list
  considers newly-established start-ups. As a matter of fact,
  incubators such as {\it 500 Startups}, {\it Y Combinator}, {\it
    Techstars} or {\it Wayra} indeed target early-stage companies,
  i.e. they make risky investments on ideas and small teams 
  without much of previous history. Their investment targets are therefore
  similar to the ones captured by our the definition of
  `open-deal list', and it is easy to realize that predicting future
  positive outcomes of firms in 
  such a set is
  more challenging than predicting future positice outcomes of more established
  firms.

\begin{figure}
\centering
\includegraphics[scale=0.5]{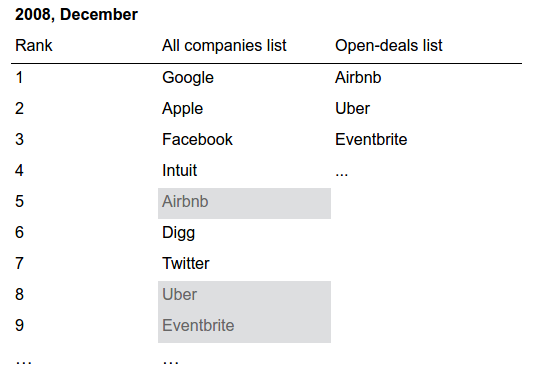}
\caption{\textbf{Example of the construction of the open-deals ranked list.} For each month of the observation period, all companies in the network were ranked according to their values of closeness centrality. Top-ranked nodes are those with the highest closeness centrality in the WWS network in the corresponding month. Only those companies (highlighted in grey) that had not yet received funding, had not yet been acquired, and had not yet been listed in the stock exchange market, were retained in the open-deals ranked list.}
\label{fig:SI:table1}
\end{figure}

Fig \ref{fig:SI:table1} shows an example of the procedure adopted. The
companies highlighted in grey are those which, prior to December 2008,
had not yet received funding, had not yet been acquired, or had no yet
been listed in any stock market. These companies thus could be seen as
investment opportunities at month $t$. Since we want to focus on
early-stage companies, we also removed any company that was more than
two years old.

\subsection{Success rate in open-deals lists}

\begin{figure}
\centering
\includegraphics[scale=0.5]{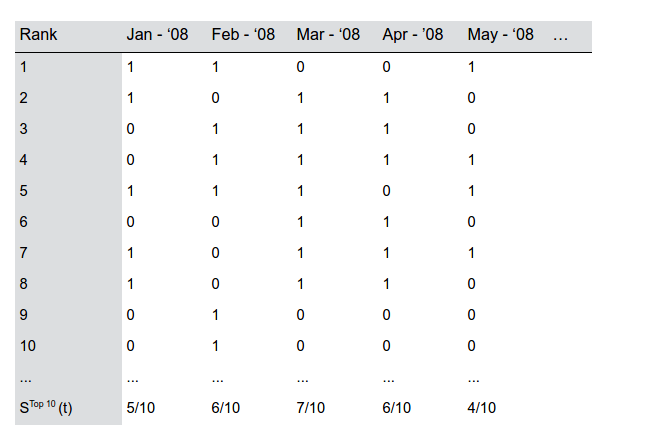}
\caption{\textbf{Illustrative example of a monthly open-deals ranking.} Companies' names are replaced by the values of their associated binary variable, with a value equal to $1$ indicating the achievement of a positive outcome. The last line reports the success rate $S^{\rm Top 10}(t)$ of companies in the Top $10$ of our recommendation list.}
\label{fig:SI:table2}
\end{figure}

Each open-deals list in month $t$ contains $M(t)$ successful companies ($0 \leq M(t) \le N(t)$), i.e., those companies that have obtained, within a time window $\Delta t = 6, 7$ or $8$ years since month $t$, a positive outcome. 
A positive outcome is here defined in terms of the occurrence of at least one of the following events: (i) the company makes an acquisition; (ii) the company is acquired; or (iii) the company undergoes an IPO. To each company in the open-deals list we then assigned the value of a binary variable, namely $1$ if the company has achieved a positive outcome within the chosen time windows $\Delta t$, or $0$ otherwise. Fig \ref{fig:SI:table2} shows an example of the monthly open-deals lists, in which names are replaced by their associated binary values. Notice that the higher the number of ones in the top regions of the rankings, the better the performance of the recommendation method in predicting positive outcomes.

\begin{figure}
\centering
\includegraphics[width=0.9\textwidth]{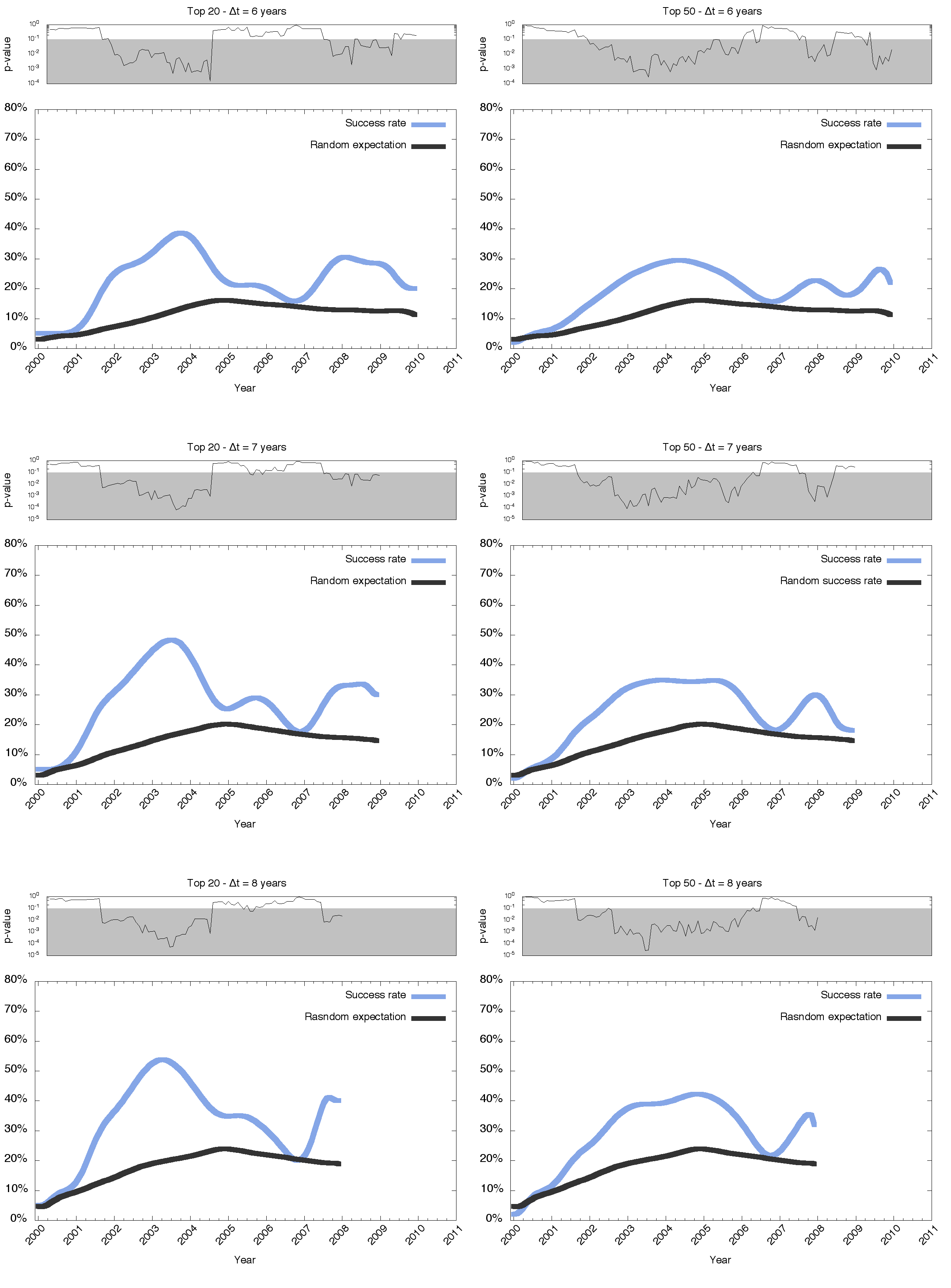}
\caption{\textbf{The success rate of our recommendation method.} The
  success rate $S(t)$ of our method (blue curve) is compared to the
  expected success rate $S^{\rm rand}(t)$ associated with the
  recommendation of randomly selected companies (black
  curve). Different lengths, namely $n=20,50$, of the recommendation
  lists, and different time windows, i.e., $\Delta t = 6,7,8$ years,
  to assess the performance of a company have been considered. The
  statistical significance of the discrepancy between $S(t)$ and
  $S^{\rm rand}(t)$ is quantified through the associated $p$-values,
  shown in the top charts of each panel.}
\label{fig:SI:success_rate}
\end{figure}

We focus on companies in the top positions of our open-deals recommendation list, and we indicate by $m(t)$ the number of companies in the Top 20 in month $t$ that have obtained a positive outcome, i.e., the number of ones in the first $n=20$ entries of the list. Notice that the same procedure has been repeated for the Top 50 companies ($n=50$) to check for the robustness of results. The accuracy of the recommendation method is assessed by computing the success rate $S(t)$ defined as the ratio $m(t)/n$. How does this compare to a null model where network properties are not taken into account? If the open-deals lists were randomly ordered, the expected number of successful companies $m^{\rm rand} (t)$ in e.g. the Top 20 ($n = 20$) would be given by the expected value of the hypergeometric distribution $\mathrm{H}(N(t),M(t),n)$. In particular, the expected value of $m^{\text{rand}} (t)$ is $nM(t)/N(t)$ and thus the expected success rate is $S^{\text{rand}}(t)=M(t) / N(t)$. Similarly, it follows that $\text{var}(S^{\text{rand}}(t))=\text{var}(m^{\text{rand}}(t))/n^2$.\\
Fig 2 (main text) and Fig \ref{fig:SI:success_rate} show that $S(t)$ (blue curve) is systematically much higher than $S^{\text{rand}}(t)$ (black curve), except during two short periods corresponding, respectively, to the dot-com bubble (1999-2001) and to the 2008 financial crisis. In both cases, the difference between $S(t)$ and $S^{\text{rand}}(t)$ becomes narrower, yet $S(t)$ always remains higher than $S^{rand}(t)$. Moreover, Fig \ref{fig:SI:success_rate} shows that these findings are robust against variations in the length of the time window (i.e., $\Delta t = 6,7,8$) and in the number of companies considered in the recommendation (i.e., Top 20 and Top 50). 

The statistical significance of the results is assessed by computing
the hypergeometric $p$-values, which give the probability of
obtaining, by chance, a success rate equal to or greater than the one
obtained with real data. Denoting as $P(\cdot)$ the probability mass function of $m^{\text{rand}}(t)$ we can compute the $p$-value at time $t$ as:
\[
p(t) =
\sum_{k=m(t)} ^ {n} P(m^{\text{rand}}(t)=k).
\]
The top charts in
Fig 2 (main text) and in each panel of Fig \ref{fig:SI:success_rate}
report the evolution of the $p$-values over time. Low $p$-values
($<0.05$) are observed in most parts of the observation period. This
suggests that the discrepancy between the success rate of the 20
top-ranked companies selected according to our recommendation method
and the success rate of the same number of companies selected at
random from the open-deals list is statistically
significant. Conversely, high $p$-values are observed in
correspondence of the downturns, thus indicating that in such periods
the success rates predicted by our recommendation method could have
been obtained also by chance.

\subsection{Real investors performance is similar to random expectation}

  It is important to highlight that, although the random
  expectation null model has mainly been introduced to assess whether
  our results are statistically significant, the performance of real
  investements is remarkably similar to the expected success rate in
  the null model. To illustrate this, a summary statistics of the
  Top 15 investors, according to the number of investments, is reported
  in Table \ref{table:realsuccess} (data extracted from
  \texttt{crunchbase.com}). Notice that there is a great variability
  in investors performance, which reflects the variability in the type
  of investments.  Highlighted in pink are those investors whose target
  complies with our definition of open-deal list. Incubators such as
  {\it 500 Startups}, {\it Y Combinator}, {\it Techstars} or {\it Wayra} focus indeed their 
  interest on very early-stage companies, i.e. they invest on ideas 
  and small teams of entrepreneurs without much history. They make the most
  risky bets in the landscape of start-ups investments and their
  performance lies around $15\%$. Their investment target is very close to  
  the type of companies that we have isolated in our definition of
  ``open deals''.
  On the other end, large venture firms such as {\it Intel Capital},
  {\it Accel Partners}, or {\it Goldman Sachs} invest in companies at later stage of
  maturity. They are interested in organizations with larger teams,
  that have already previously received funding, and they typically
  inject funds to boost a business that has already found a market fit
  and has history of revenues, customers, and other indicators of growth.
  The presence of quantitative indicators of growth allows large venture firms
  to perform a more objective evaluation of the company and its success
  potential, which in turn is reflected on higher investment
  performances.
  
   \begin{table}[h]
\small{
\begin{tabular}{|c|c|c|c|}
\hline
{\bf Investor}&{\bf \# investments}&{\bf \# successful investments}&{\bf Success rate}\\
\hline
\rowcolor{Lavender} {500 Startups} &1022 &153 & ${\bf 15\%}$ \\
\rowcolor{Lavender}{Y Combinator} & 953 & 154& ${\bf 16\%}$\\
Intel Capital & 744 & 313 & $40\%$\\
Start-up Chile & 710 & 10 & $1.4\%$\\
Sequoia Capital & 700& 267 & $38\%$\\
New Enterprise Associates (NEA) & 672 & 272 & $40\%$\\
SV Angel & 600 & 258 & $43\%$\\
\rowcolor{Lavender}{Techstars} & 549 & 95 & ${\bf 17\%}$\\
Brand Capital & 537 & 80 & $14\%$\\
Accel Partners (Accel) & 536 & 270 & $50\%$\\
Sos Ventures (SOSV) & 493 & 17 & $3\%$\\
\rowcolor{Lavender}{Wayra} & 476 & 11 & ${\bf 2\%}$\\
Kleiner Perkins Caufield \& Byers (KPCB) & 457 & 203 & $43\%$\\
Right Side Capital Management (RSCM) & 449 & 44 & $10\%$\\
Goldman Sachs & 410 & 209 & $50\%$\\
\hline
\end{tabular}
\caption{Top 15 investment companies
    according to the number of investments made, along with the
    percentage of successful investments. Highlighted in pink are investors focused on very
    early-stage companies as those considered in our open-deal
    lists. The success rates of such investors are comparable to 
    the random expectation null model, and much below the success
    rate obtained using our recommendation method.}} \label{table:realsuccess}
\end{table}

\noindent In summary, while investors decide on which
  start-ups to invest through costly and labour-intensive screening
  processes, results confirm that the percentage of real investments
  that were deemed `successful' is consistently similar to the success
  rate given by our random expectation model. In other words,
  state-of-the-art success rate is not much better than a random
  expectation null model. This means that any improvement upon the
  null model provides valuable information. We conclude that our
  recommendation method based on centrality --whose success rate
  consistently exceeds random expectation over several periods-- is a
  considerable improvement with respect to the state of the art.

\subsection{Details on overall success rate}

To obtain an overall measure of the performance of our method, the
success rate can be aggregated across the entire observation
period. This can be carried out in two complementary ways leading to
two different measures of the overall success, namely
$\widetilde{S}_\text{I}$ and $\widetilde{S}_{\text{II}}$. Here we
discuss and provide some details with regards to both measures.

\begin{figure}
\centering
\includegraphics[width=0.8\textwidth]{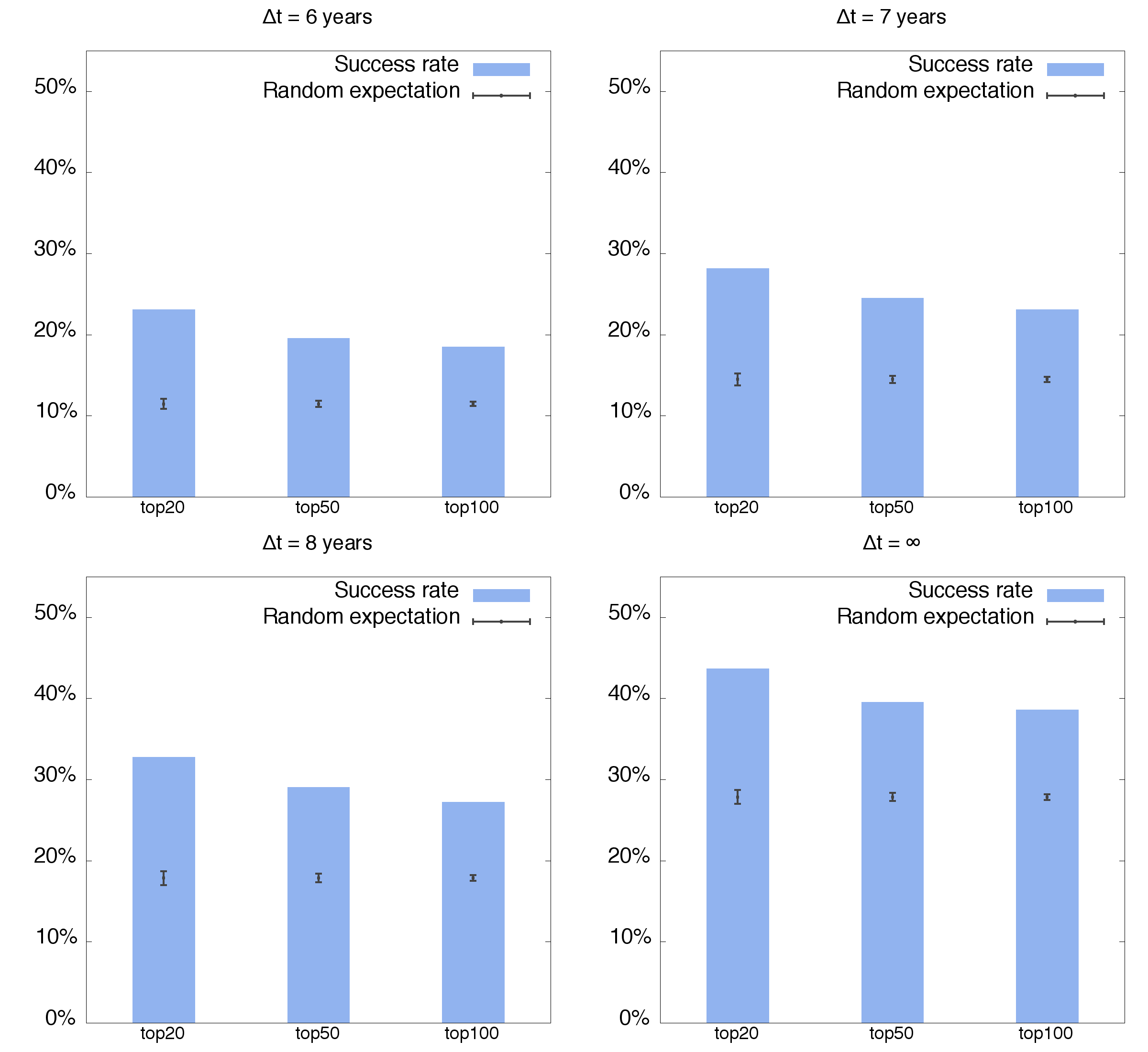}
\caption{\textbf{Observed and randomly expected success rates}. The overall success rate empirically found $\widetilde{S}_\text{I}$ (blue bars), the overall success rate $\widetilde{S}  ^{\rm rand} _\text{I}$ expected by chance (black dots), and its standard deviation (black error bars), for various values of $\Delta t$ and lengths of the list of top-ranked companies (i.e., Top 20, 50, and 100).}
\label{fig:SI:overall_success}
\end{figure}

The first measure of overall success rate, $\widetilde{S}_\text{I}$, which is used in the main text, takes into account the total number of positive entries in the top positions in all open-deals lists, regardless of the specific companies that occupy those positions. In this way $\widetilde{S}_\text{I}$ provides a measure of the overall goodness of the ranking across months, but it does not provide information about the number of unique companies correctly or wrongly identified as successful. As an example of the computation of $\widetilde{S}_\text{I}$, let us consider the period starting in January 2000 and ending in December 2007, and the Top 20 companies (bottom-left charts in Fig \ref{fig:SI:success_rate}). Such a period includes $\delta = 96$ months. The overall success rate $\widetilde{S}_\text{I}$ is defined as: 
$$ \widetilde{S}_\text{I} = \frac{   \widetilde{m}_\text{I}   }{  \widetilde{n}_{\text{I}}  }, $$
\noindent where $\widetilde{n}_{\text{I}} =  20 * \delta$ is the total number of entries in the Top 20 list across the $\delta$ months, and $\widetilde{m}_\text{I} = \sum_{t} m(t)$, where the sum runs over all months in the observation period. To construct a null model to which we can compare these measures,   we then proceed to randomly shuffling the entries in each open-deal list independently for each month and apply the same procedure (i.e., the null model makes a random sampling of the list without replacement). Accordingly, at month $t$ we count the number of successful companies within the Top 20 and label it $m^{\rm rand} (t)$. The expected total number of successful companies within all the Top 20 lists in this null model is thus given by:
$$ \widetilde{m} ^{\rm rand} _\text{I} = \sum_t  m^{\rm rand} (t), $$
\noindent and the corresponding variance is given by the sum of the variances in each month
$$\text{var}( \widetilde{m}^{\rm rand} _\text{I}) = \sum_t  \text{var}( m ^ {\rm rand} (t)), $$
where $\text{var}( m ^ {\rm rand} (t))$ denotes the variance associated to the random null model, i.e. the variance of the hypergeometric distribution.
\noindent The expected overall success rate in the case of random ordering is then given by
$$ \widetilde{S}^{\rm rand} _\text{I} = \frac{\widetilde{m} ^{\rm rand} _\text{I}} {\widetilde{n}_{\text{I}}}, $$
\noindent and its standard deviation $\sigma_\text{I}$ is
$$ \sigma_\text{I} = \frac{ \sqrt{ \text{var}( \widetilde{m}^{\rm rand} _\text{I} )}} {\widetilde{n}_{\text{I}}}. $$
\noindent Figure \ref{fig:SI:overall_success} reports the overall success rate empirically found $\widetilde{S}_\text{I}$ (blue bars), the overall success rate $\widetilde{S} ^{\rm rand} _\text{I}$ expected by chance (black dots), and its standard deviation (black error bars) for various values of $\Delta t$, and for different numbers of recommended companies (i.e., Top 20, 50, and 100). 

The second measure of the overall success rate, $\widetilde{S}_{\text{II}}$, does not simply capture the overall performance of the ranking-based recommendation method, but compares the number of unique companies in the Top 20s correctly predicted as successful by our method, across the entire observed period, against the number of successful companies that would be expected under random selection. In particular, this second measure of overall success is based on: (i) the total number $\widetilde{N}_{\text{II}}$ of unique companies available in any month; (ii) the total number $\widetilde{M}_{\text{II}}$ of unique companies that have achieved a positive outcome at any time since their foundation up to 2015; (iii) the number $\widetilde{n}_{\text{II}}$ of unique companies included in all Top 20 rankings in any month; and (iv) the number $\widetilde{m}_{\text{II}}$ of unique companies, listed in all Top 20 rankings, that have achieved a positive outcome at any time since their foundation up to 2015. 

\begin{figure}
\centering
\includegraphics[width=0.5\textwidth]{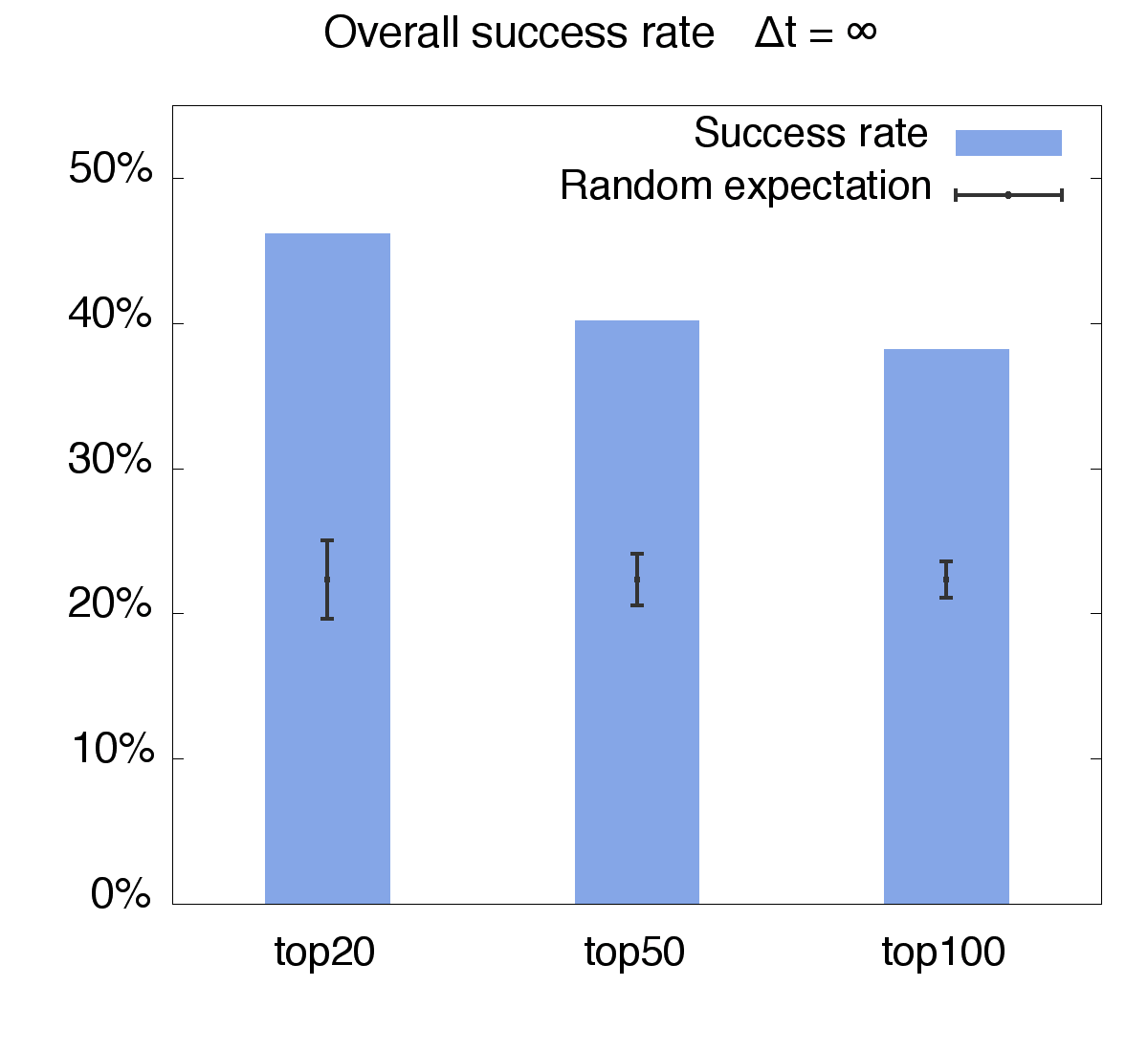}
\caption{\textbf{Observed and randomly expected success rates.} The overall success rate $\widetilde{S}_{\text{II}}$ (blue bars) obtained through the second method of aggregation assessed against the overall success rate $\widetilde{S}  ^{\rm rand}_{\text{II}}$ expected by chance (black dots), and its standard deviation (black error bars), for various lengths of the list of top-ranked companies (i.e., Top 20, 50, and 100).}
\label{fig:SI:overall_success_prima_o_poi}
\end{figure}

Notice that, in this way, each company contributes only once to the evaluation of the success rate. Therefore, the probability of finding exactly $\widetilde{m}_{\text{II}}$ successful companies in any ranking of Top 20 (50, or 100) is given by the hypergeometric function $\mathrm{H}( \widetilde{N}_{\text{II}}, \widetilde{M}_{\text{II}}, \widetilde{n}_{\text{II}}, \widetilde{m}_{\text{II}})$. The success rate shown in Fig \ref{fig:SI:overall_success_prima_o_poi} is computed as $\widetilde{S}_{\text{II}} = \widetilde{m}_{\text{II}} / \widetilde{n}_{\text{II}}$, while the success rate $\widetilde{S}_{\text{II}}^{\rm rand}$ in the case of the null model is given by $\widetilde{S}_{\text{II}}^{\rm rand} =  (\widetilde{M}_{\text{II}} / \widetilde{N}_{\text{II}})$. Fig \ref{fig:SI:overall_success_prima_o_poi} reports also the error bars of the success rate computed as the standard deviation of the hypergeometric distribution.

While the first index of overall performance assesses the average goodness of the ranking, the second index measures only the number of companies correctly identified as successful across the entire observation period. The two aggregation methods produce comparable results, and achieve a substantial success rate of about $40\%$ in the case of $\Delta t = \infty$. Moreover, in both cases, the success rate found in reality and the one expected by random chance are very different, and their discrepancy is always statistically significant with $p$-values smaller than $10^{-5}$.

\section{Additional analysis}

\subsection{Closeness centrality in 
successful vs non-successful start-ups}

\begin{figure}
\centering
\includegraphics[width=0.55\textwidth]{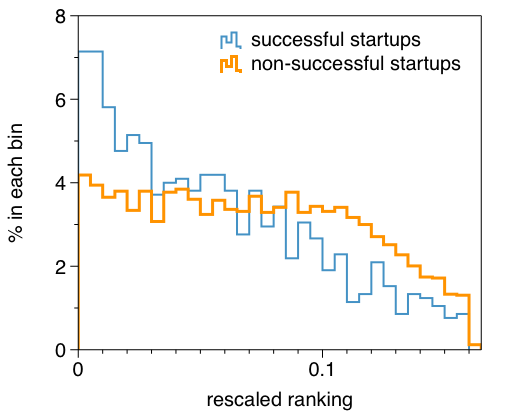}
\caption{\textbf{Closeness centrality distributions.} 
Histograms of closeness centrality rescaled rankings (in bins of 0.005, see the text) for
start-ups which will have a positive outcome (blue) and for which no positive outcome occurs (orange).
Successful start-ups have statistically lower rescaled rankings 
(i.e. higher centralities) than non-successful ones. For every
start-up, only the value of closeness centrality collected in the last month of
observation has been used. }
\label{fig:SI_ranking_histograms}
\end{figure}

To have a better understanding of how closeness
  centrality is distributed among start-ups, in Fig
  \ref{fig:SI_ranking_histograms} we compare the estimated frequency
  histograms of closeness centralities rescaled ranking. To
  obtain the rescaled ranking, in each calendar month we
  calculated the closeness centrality of each firm in the global
  network and ranked all firms in terms of their centrality, what gives an `absolute rank` for each firm. We then extract those firms which belong to the open-deal list, and re-rank them accordingly (so that the firm with top ranking acquires a rank $0$ in the open-deal ranking, the second acquires rank $1$, and so on). The rescaled ranking is then defined as the ratio between the open-deal-rank and the maximum absolute-rank of open-deal companies at a given month.
  Thus, the firm with the highest
  position (i.e., zero ranking) maintained the same value (i.e., zero)
  in the rescaled ranking. By contrast, firms at lower positions were
  assigned rescaled values approaching 1 as their ranking approached
  the highest value (i.e., the lowest position). Such a rescaling 
  thus enables to appropriately compare 
  firms characterised by different values of
  centrality, obtained in different networks and at different calendar
  times. In order to smooth out the data, a binning has been performed in the x axis (bin size of $0.005$).
%
  We notice
that histograms are non-overlapping, and that there is a net overabundance of start-ups with a positive outcome (successful) closer to the top of the ranking. In other words, start-ups which are higher in the centrality rankings
(i.e. small values of closeness centrality ranks) have statistically a higher
chance of positive economic outcomes. This confirms that rankings
based on closeness centrality are indeed informative of a start-up long-term success
and can then be used to inform recommendation.

\subsection{Different centrality measures are correlated}

We have considered closeness centrality as our primary
  measure of network centrality. Closeness centrality is based on the lengths
  of shortest paths in the network. However, the structural centrality of a node
  in a network can be quantified by different network metrics, either global such
  as closeness and betweenness, and local as the degree centrality \cite{Wasserman1994}.
  {\it Closeness centrality} of a node (Eq.\ref{def_closeness}) characterises the overall distance between that node and the rest of the nodes in the network, such that the lower that overall distance, the higher this measure, and hence the more central this node is.\\  
  On the other hand, the {\it betweenness centrality} $b_i(t)$ of a node $i$ when the network is observed at a given time $t$ is given by
  \begin{equation}
  b_i(t)=\frac{1}{(N-1)(N-2)}\sum_{j=1,j\ne i}^N \sum_{k=1,k\ne i,j}^N \frac{n_{jk}(i; t)}{n_{jk}(t)},
  \label{def_betweenness}
  \end{equation}
where $n_{jk}(t)$ is the total number of shortest paths between nodes $j$ and $k$ whereas $n_{jk}(i)$ is the number of shortest paths between $j$ and $k$ that actually go through $i$.   This measure was introduced by Freeman to quantify the fact that communication travels {\it just} along shortest paths, and so a node $i$ is more `central' the more shortest paths among pairs of nodes in the network go through it.\\
While both closeness and betweenness are measures of centrality based on shortest paths, one can also think of a node being central if it acquires many edges over time --i.e. acquiring intel from several other companies--. To account for this we may resort to use {\it (normalised) degree centrality} $d(i)$, defined as
\begin{equation}
d_i(t)=\frac{k_i(t)}{k_{max}(t)},
\end{equation}
where $k_i(t)$ is the degree (number of links) of node $i$ and $k_{max}(t)$ is the largest degree in the network at that particular time snapshot.\\

\begin{figure}
\centering
\includegraphics[width=0.99\textwidth]{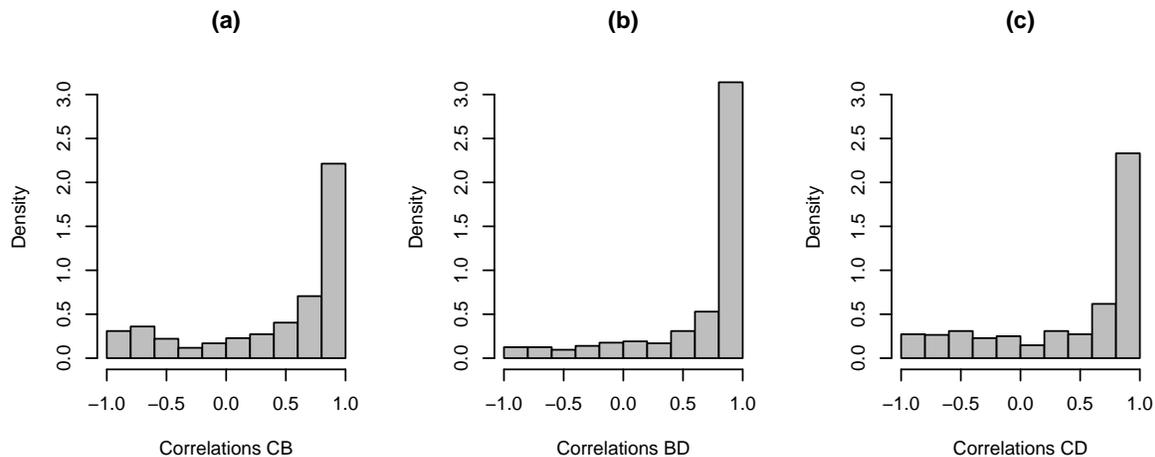}
\caption{\textbf{Centrality measures are correlated. }
  Frequency histograms of the Pearson correlation
    coefficients between (a) rescaled closeness and rescaled betweeness centrality,
    (b) rescaled betweeness and rescaled degree centrality, and (c) rescaled closeness and
    rescaled degree centrality. We find that centrality measures
    are systematically correlated between each other, so the choice of using
    closeness centrality as the centrality measure under analysis is robust.
     }
\label{fig:SI_correlations}
\end{figure}

Consequently, the centrality of a start-up
  in the WWS network can be measured in many
  alternative ways. In this section we will show that the choice of using
closeness centrality is not only supported by theoretical arguments based on
employees' mobility and intel flows among companies, but it also a robust
choice as other alternative measures produce similar results. 
\\
To validate robustness, for each start-up in the open-deal list across
time we have computed additional centrality measures, namely degree
and betweeness centrality \cite{bookCN}, and computed to which extent
all three possible measures of centrality are correlated. More
concretely, we consider all start-ups in the open-deal list for which (i) we
have data of the three centralities over at least 3 of the 24 months
forming the observation window, and for which (ii) closeness and betweeness centralities are defined. For each firm, we then compute the
Pearson correlation coefficient between the monthly sequence of each
pair of (rescaled) centrality measures. We do this for all firms and we then construct
the frequency histogram of the Pearson correlation coefficients. Results
are reported in Fig. \ref{fig:SI_correlations}. Interestingly, we find
that the three measures are in general well (pairwise)
correlated. 
%
We conclude that the choice of a particular type of global centrality
measure, such as closenness, is a robust choice as other global
structural indicators based on a different use of shortest paths and,
to a minor extent, also local measures such as the degree are
correlated with the closeness in the case of the WWS network under
analysis in this work.  Hence, focusing on closeness centrality is a
robust choice. In the next subsection we round-off this validation by
exploring results of our recommendation method using either
betweenness or degree centrality as the key network indicator, and
will show that success rates of the recommendation method are similar
in all three cases.

\subsection{Recommendation methods based on other
    centrality measures}
\begin{figure}
\centering
\includegraphics[width=0.85\textwidth]{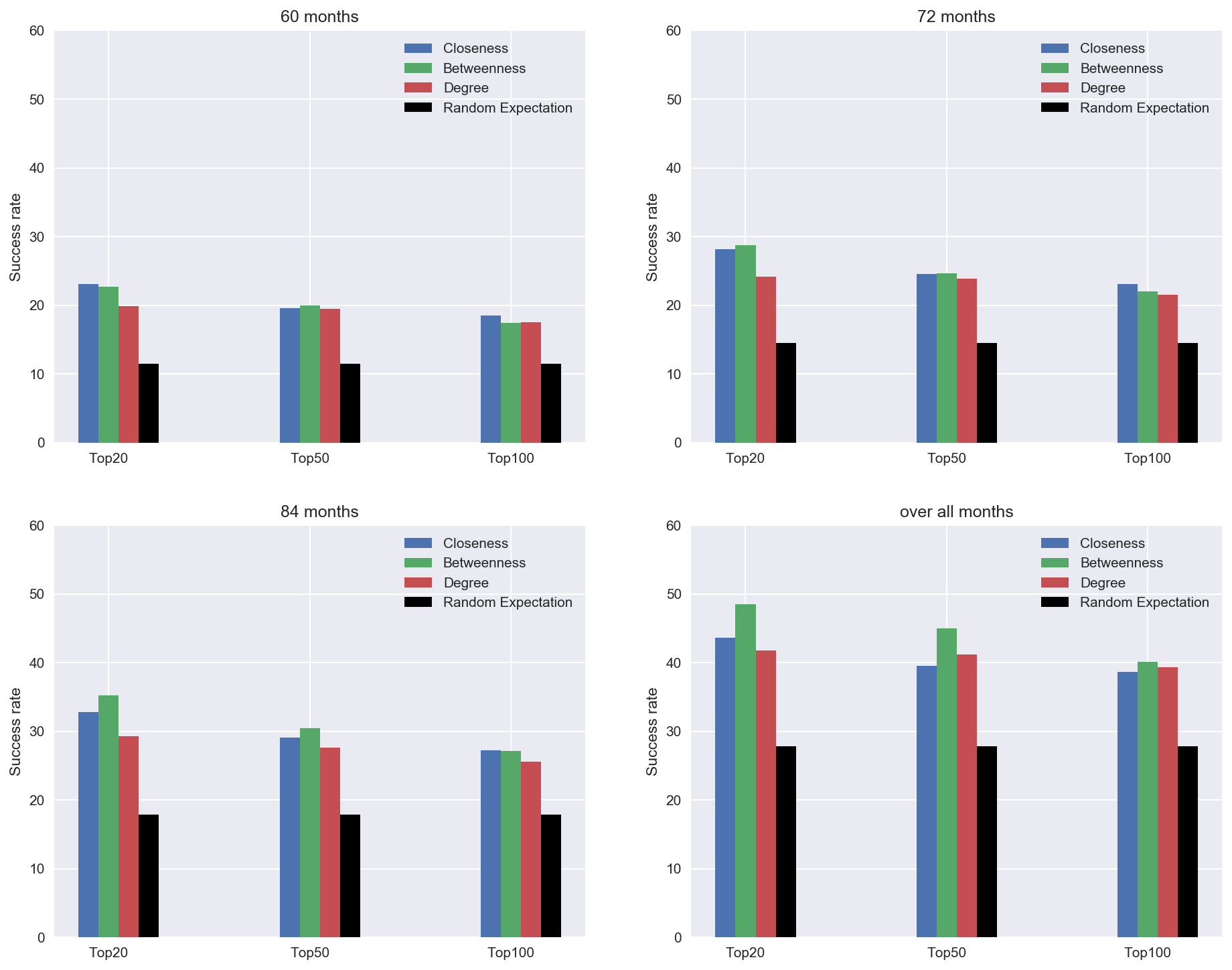}
\caption{\textbf{Overall success rate of recommendation methods based
    on different centrality measures. } Comparison of
    recommendation methods focusing on the top 20, top 50 and Top 100
    rankings, for $\Delta t=5, 6, 7$ and $\infty$ years, using
    different centrality measures.  Closeness, degree and betweeness
    centralities perform similarly, and recommendations based on
    either of these measures are systematically superior to a random
    expectation model, with overall success rates which are systematically larger than in the null model.}
\label{fig:allcentralities}
\end{figure}

\begin{figure}
\centering
\includegraphics[width=0.65\textwidth]{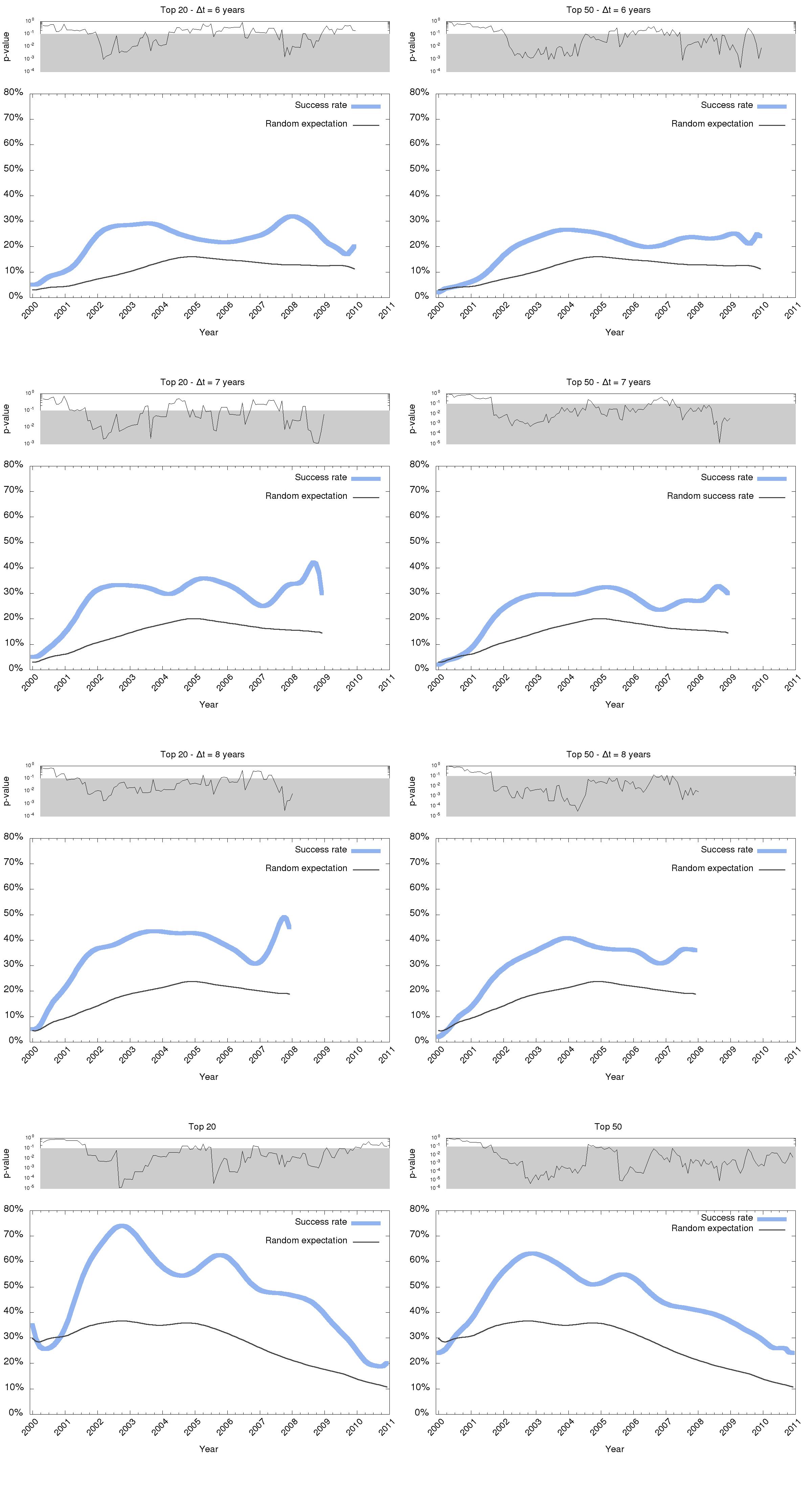}
\caption{\textbf{Monthly success rate of recommendation methods based on betweeness centrality. }
  Recommendation methods
  focusing on the Top 20 and Top 50 ranked start-ups in the open-deal
  list, for $\Delta t=5, 6, 7$ and $\infty$ years, using betweeness
  centrality instead of closeness centrality. Results are
  qualitatively similar and the monthly success rate of  
  recommendations based on betweenness 
  is systematically superior to that of a random
  expectation model.}
\label{fig:allcentralities2}
\end{figure}

\begin{figure}
\centering
\includegraphics[width=0.65\textwidth]{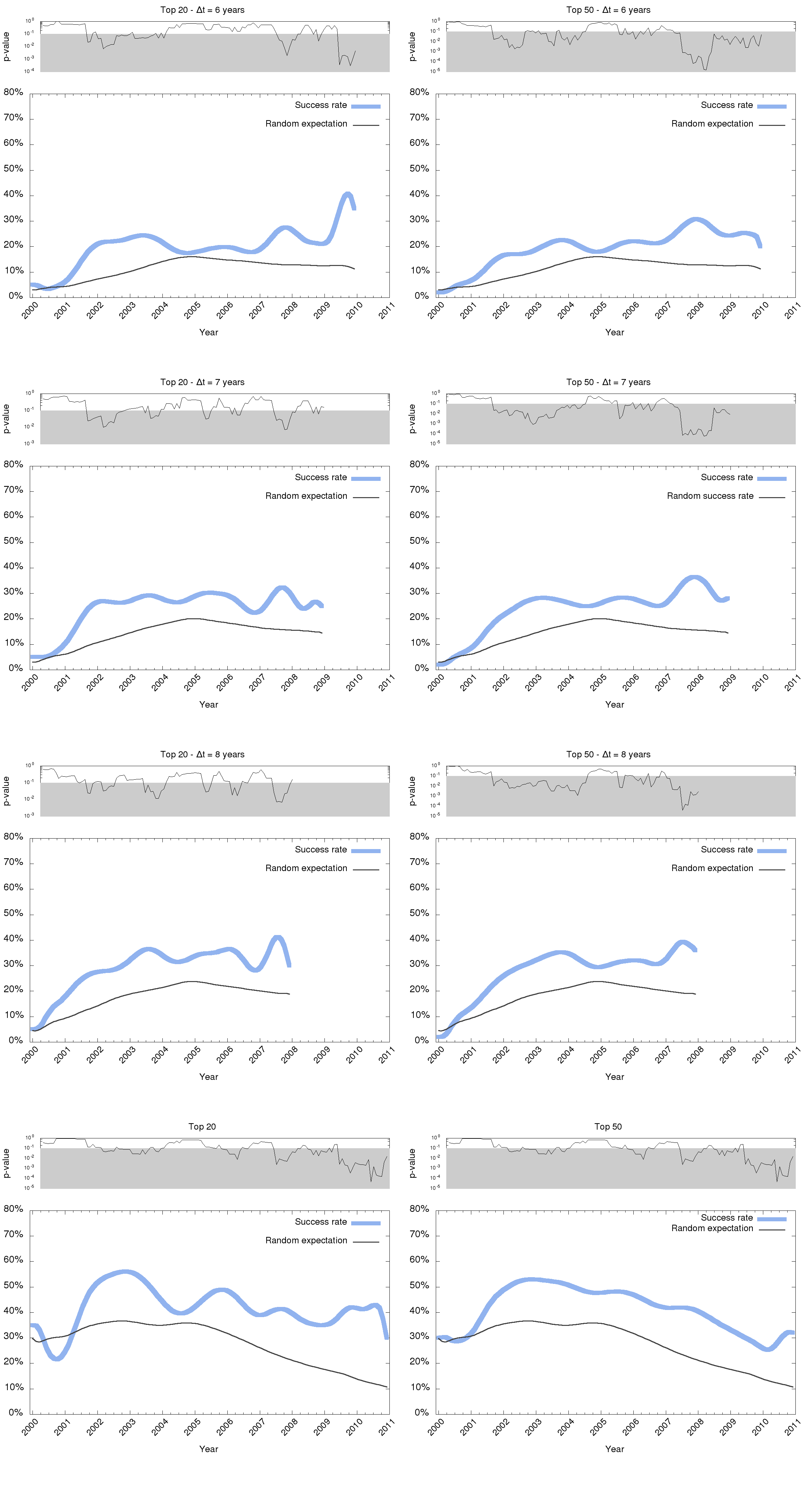}
\caption{\textbf{Monthly success rate of recommendation methods
    based on degree centrality. }
  Recommendation methods
    focusing on the Top 20 and Top 50 ranked start-ups in the open-deal
   list, for $\Delta t=5, 6, 7$ and $\infty$ years, using degree
   centrality instead of closeness centrality. Results are
   qualitatively similar and the monthly success rate of 
  recommendations based on degree 
   centrality is systematically superior to that of a random
   expectation model.}
\label{fig:allcentralities3}
\end{figure}

To further complement the correlation analysis of the
  previous subsection, here we focus on recommendation methods based
  on centrality measures other than closeness. Results are summarised in
  Fig.\ref{fig:allcentralities} for averages over the entire period, and in
  Figs.\ref{fig:allcentralities2} and \ref{fig:allcentralities3} for
  monthly analysis. In every case we find that the results are
  qualitatively similar whether we use closeness, betweeness or degree 
  centrality, with success rates systematically larger than random
  expectations (and therefore larger than the actual perfomance of accelerators and investors focusing on early-stage start-ups).

\subsection{The effect of fading links}
The mobility of workers from one company to another
  creates an intel flow between companies. Our working hypothesis is
  that companies receiving employees increase their fitness by
  capitalising on the know-how the employee is bringing with him/her.
  Such microscopic dynamics is thus captured and modelled by the
  creation of new edges at the level of the network of start-ups. As a
  consequence, companies which are perceived at the micro scale as
  appealing opportunities by mobile employees will likely boost their
  connectivity and therefore will acquire a more central position in
  the WWS network. An important underlying assumption is that, once a
  link is created, it will remain in the network indefinitely, so that   
  the company that has received the intel keeps it and builds on this intel
  forever.  
    Conversely, considering the
  possibility of removing links (or actually fading their strength) some time after their creation,  
  would actually be equivalent to assume that companies 
  can lose the know-how they have acquired, something which is less
  likely to occur.
Accordingly, allowing links to fade or be removed with time in the construction of
the time-varying WWS network should lead to recommendations on the positive
economic outcome with much lower success rates than those obtained
from a network where know-how is not artificially removed. To check
for this case, we have first build the WWS (for each month) from January 1990 to December 1999. Then, starting from January 2000 onwards, for each month all connections older than 10 years are removed from the network. Closeness is then evaluated each month as described in the recommendation method. A similar analysis is also performed for 5-year fading instead of 10-year fading, with very similar results.\\

\begin{figure}
\centering
\includegraphics[width=0.85\textwidth]{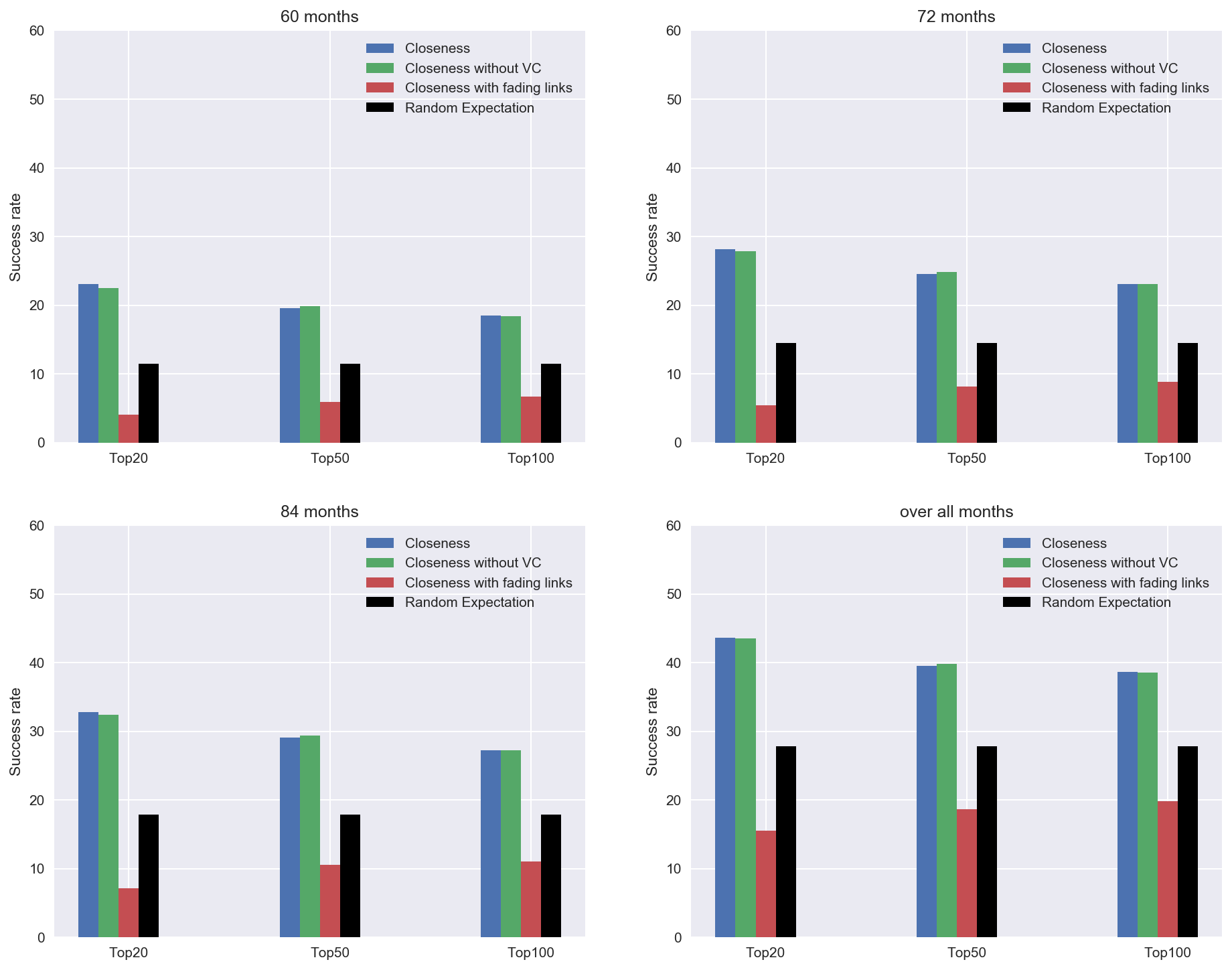}
\caption{\textbf{Effects of fading links and removal of venture
    capital funds.} 
  Success rate of recommendation
    methods focusing on the Top 20, Top 50 and Top 100 rankings, for
    $\Delta t=5, 6, 7$ and $\infty$ years. The standard case of closeness
    centrality from the original network (blue bars) is compared to  
    closeness centrality in a case where the links of the WWS are allowed
    to fade over time (red bars), and to closeness centrality 
     in a situation where all
    venture capital funds (VC) have been removed from the WWS network (green bars). Results from the null model are plotted in black bars. 
}
\label{fig:VC_removed}
\end{figure}

\begin{figure}
\centering
\includegraphics[width=0.65\textwidth]{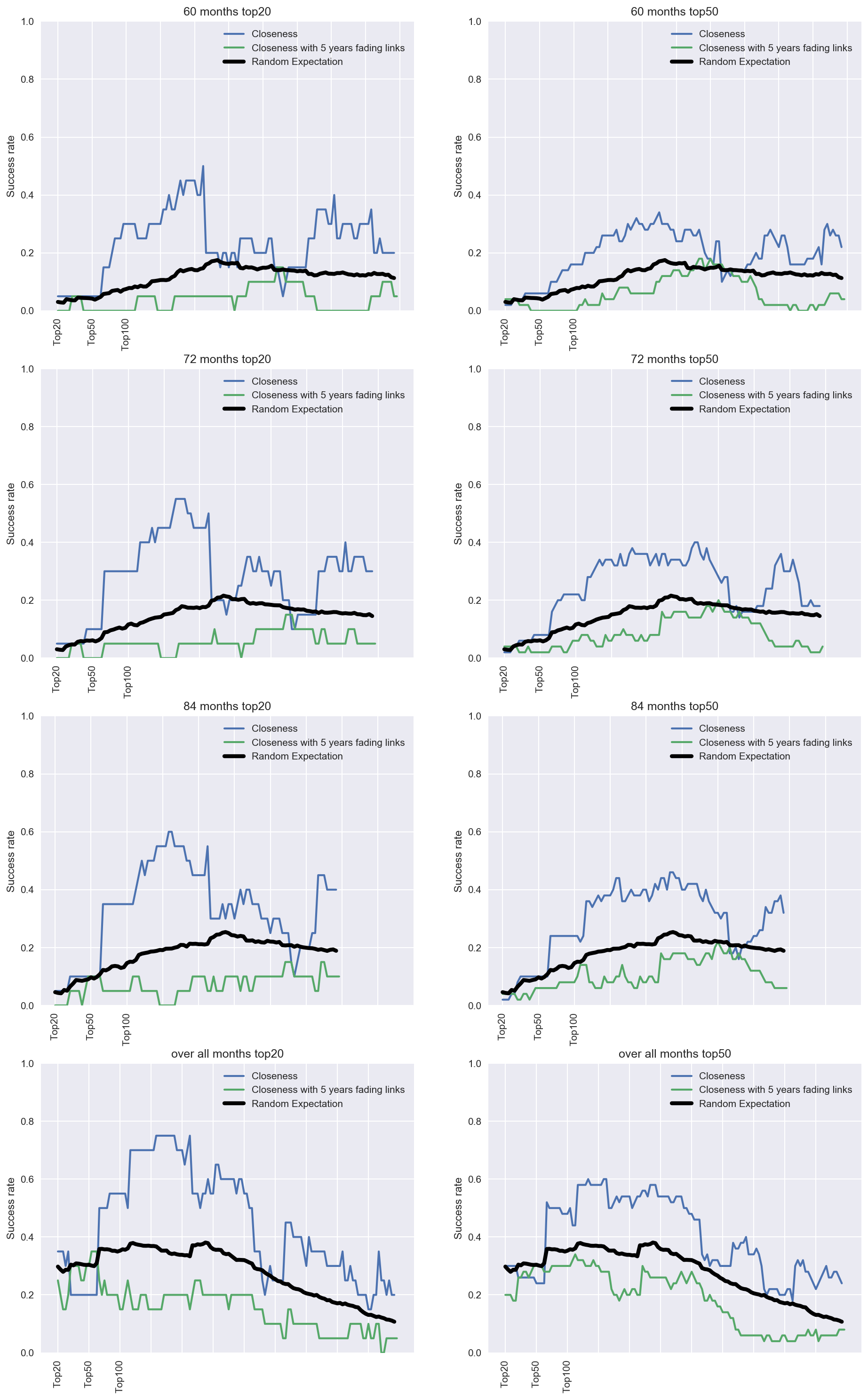}
\caption{\textbf{Effect of fading links on monthly success rates. }
  Systematic comparison of the monthly success rate of
    a recommendation based on the closeness centrality
    from the original network and in a case where the links of the WWS are allowed
    to fade over time. In the latter case success
    rates drop below the random null model.}
\label{fig:fading}
\end{figure}

\noindent 
In Fig.\ref{fig:VC_removed} we compare the overall success rate for the 5-year fading case
(red bars) to our standard recommendation method based on a
WWS network that does not allow link fading. Results show that a
recommendation method with fading links systematically fails. In fact it
works even worse than a random null model, in good logical agreement with our previous discussion. For completeness, a
comparison of the two methods is also considered for 
the monthly success rates in Fig.\ref{fig:fading}. Results are consistent
with those obtained for the overall success rate.

All these results strengthen our working hypothesis that the intel
flow across start-ups is well captured by node centrality in the WWS
network.

\begin{figure}
\centering
\includegraphics[width=0.65\textwidth]{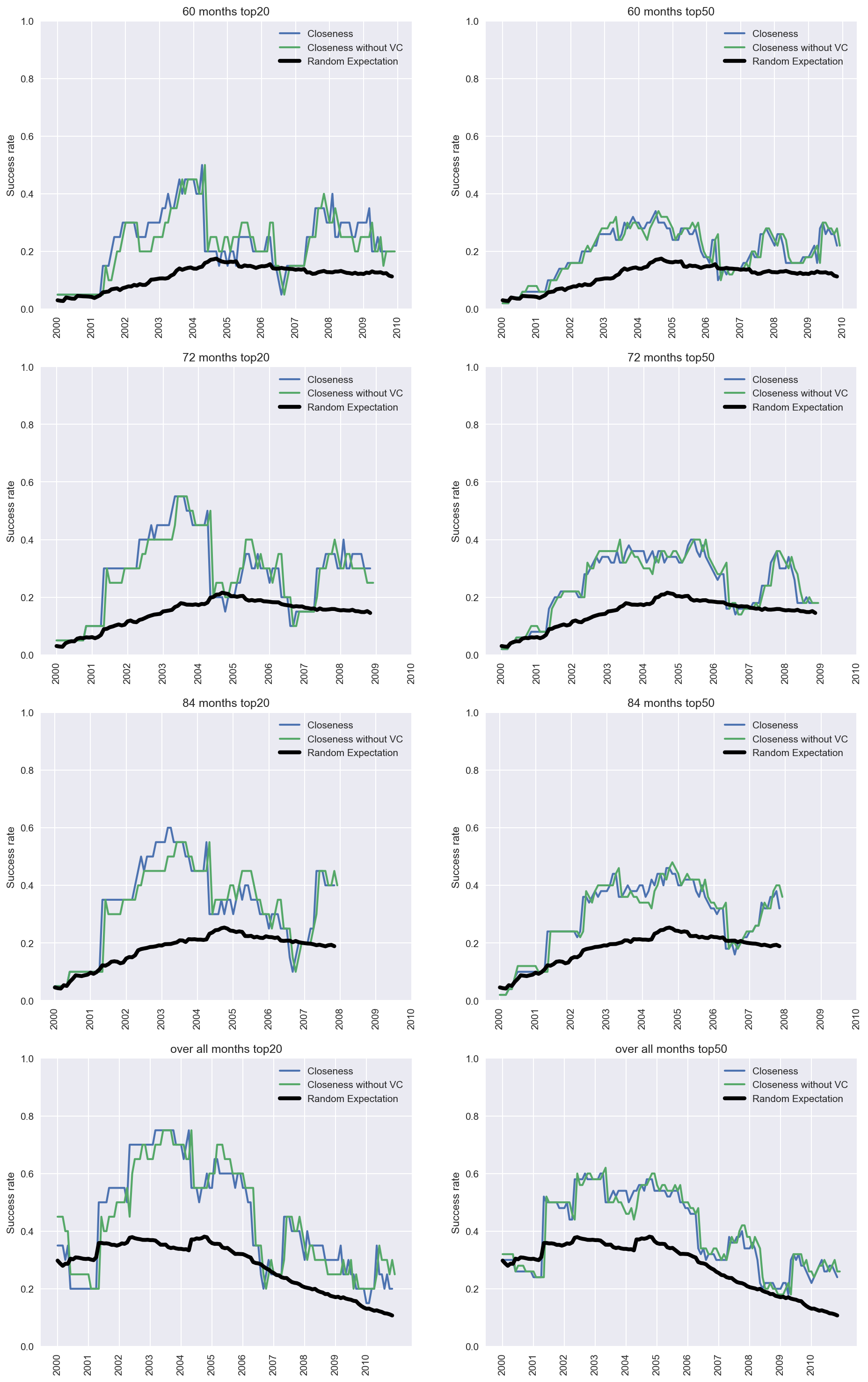}
\caption{\textbf{Comparison of monthly success rate of recommendation method based on a worldwide start-up network with and without venture capital funds. }
  Monthly success rate of the recommendation method focusing in the Top 20 and Top 50, for $\Delta t=5, 6, 7$ and $\infty$ years, comparing the original case to the case where all venture capital funds have been removed from the worldwide start-up network. Results are essentially identical to the ones obtained in the original case, hence confirming that the topological presence of venture capital funds is not a confounding factor.}
\label{fig:closeness_VC}
\end{figure}

\subsection{Possible confounding factor 1: the effect of venture capital funds}

A first possible confounding factor is the presence of venture capital
funds, i.e. the fact that the presence of these nodes in the network
might enhance the closeness centrality of start-ups.  In order to
assess the role played by venture capital funds in the effective
centrality of different start-ups, we have performed an experiment
where we remove all venture capital funds from the world start-up
network, and subsequently have recomputed closeness centrality values
for each start-up in the open-deal list. Concretely, we extracted from
\texttt{CrunchBase.com} a list of $101$ companies that are labelled as
venture capital firms see Table \ref{table:VC} for details.
\\
\noindent Accordingly, in this experiment we create the WWS network but not include those nodes in the network (and all the connections they bring with them). Closeness centrality is then evaluated each month as described in the recommendation method.
\noindent Results of overall success rate are shown (green bars) in Fig.\ref{fig:VC_removed} while monthly success rates are compared in Fig.\ref{fig:closeness_VC}. The success rate of the recommendation method based on this quantity is consistently similar to the one found in the case where venture capital funds are not removed from the original network, hence confirming that the topological presence of venture capital funds is not a confounding factor.

\subsection{Possible confounding factor 2: number of employees}
\begin{figure}
\centering
\includegraphics[width=0.43\textwidth]{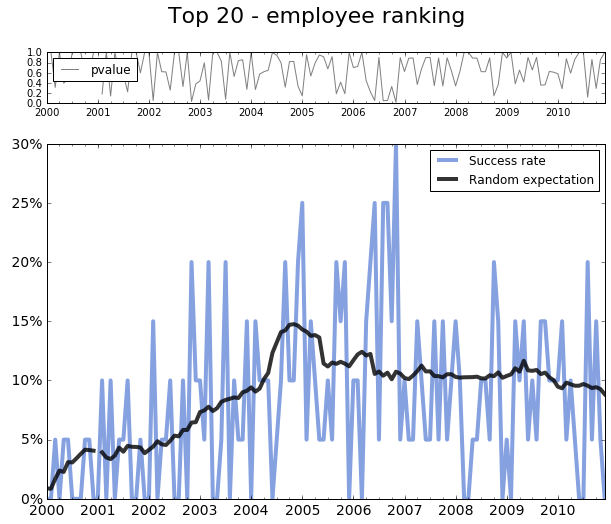}
\includegraphics[width=0.45\textwidth]{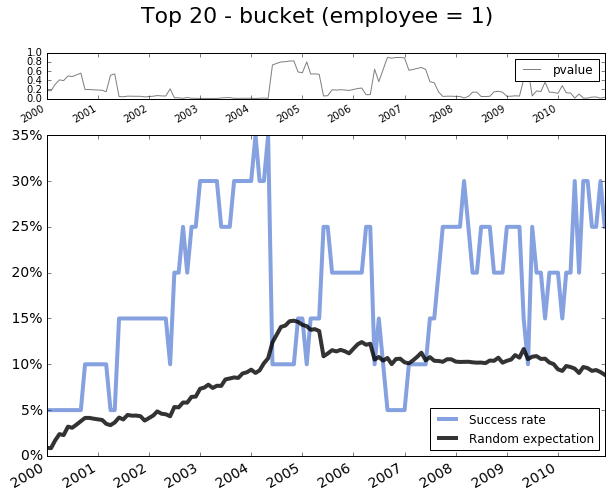}
\caption{\textbf{Accounting for start-up size. } (Left panel) Monthly success rate of an hypothethical recommendation method {\it based only on start-up size} (number of employees), focusing in the Top 20 start-ups in the open-deal list. Results wildly fluctuate above and below a random expectation model, and p-values safely conclude that the number of employees is not informative of the start-up success likelihood. (Right panel) Monthly success rate of the recommendation method based on closeness centrality, focusing in the Top 20 start-ups from a subset of the open-deal list {\it gathering only start-ups with a single employee}. Results are qualitatively similar to the ones obtained without conditioning for start-up size, and suggest that start-up size is not a confounding factor.}
\label{fig:success_employee}
\end{figure}

A second possible confounding factor or hidden predictor is the start-up size (e.g., number of employees). To assess this possibility, we have conducted a number of experiments. Initially, we explored start-up size (number of employees) instead of topological network centrality as the informative predictor, and built a recommendation method based on that metric. Results are shown in the left panel of Fig.\ref{fig:success_employee}, confirming that number of employees is not informative of the start-up success likelihood.\\
Additionally, we have also checked the recommendation method (based on closeness centrality) when only the subset of open-deal start-ups with a fixed number of employees is considered. Since the most frequent size is a start-up with a single employee, we extract the subset of all start-ups with only one employee. Monthly success rates of the recommendation method are shown in the right panel of Fig.\ref{fig:success_employee}. These results confirm that start-up size is not a counfounding factor and that number of employees is indeed not an informative variable that determines future success.

\subsection{Possible confounding factor 3: start-up geographical
    location}
\begin{figure}
\centering
\includegraphics[width=0.65\textwidth]{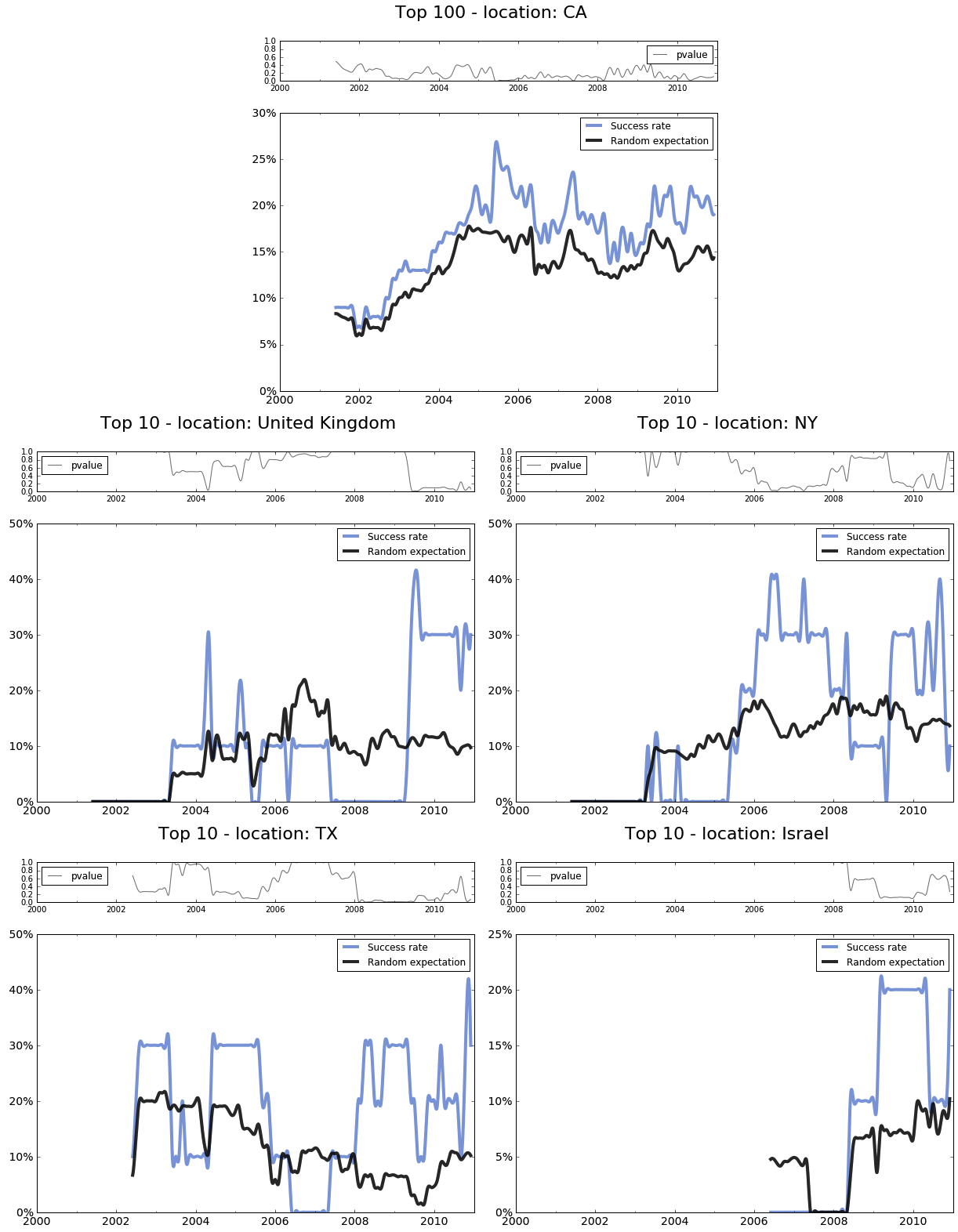}
\caption{\textbf{Accounting for spatial location.} Monthly success rate of recommendation method on five geographically separated regions.: California, United Kingdom, New York (US), Texas (US) and Israel. For the California ecosystem we considered the Top100 ranking whereas for the other four (smaller) ecosystems we only considered the Top 10.
Results are more noisy by qualitatively similar to the ones obtained in the original case, hence confirming that spatial location is not a confounding factor.}
\label{fig:locations}
\end{figure}

A third possible confounding factor is the  geographical location of each of the start-ups. To account for this, we have replicated our analysis (originally performed at a worldwide scale) in five geographically separated regions, by dividing open-deal start-ups in five subsets: California, United Kingdom, New York, Texas and Israel. Results for the monthly success rate of our recommendation method are plotted in Fig.\ref{fig:locations}. While results are more noisy than for the worldwide setup, we can confirm that for every case the recommendation method based on closeness centrality is above the random expectation.

\section{From recommendation to prediction of start-up success:
    supervised learning approaches}

The recommendation method proposed in the main text is based on the
working hypothesis that start-ups with higher closeness centrality
rankings are more likely to experience a economic successful outcome
in the future. We have provided theoretical foundation to our research
hypothesis at the microscopic level, and then heuristically validated
our recommendation lists on a monthly basis, obtaining results that
are significantly better than those obtained by a random expectation
model.
\\ 
However, strictly speaking, a recommendation method is not a true prediction
method, as we are not predicting the outcome of each and every start-up in the
open-deal list (either to the successful or to the non-successful category).
To bridge this gap, in this section we consider different types of prediction models which can
indeed truly ``predict'' the positive outcome of a start-up, i.e. they can
classify whether a given start-up will have a positive outcome or not.
\\
All models are initially based on a sample including $5,305$ firms.
These are the firms that have been in the open-deal list for at least
one month, and can therefore be suggested as potential investment
opportunities. Each firm is then observed over a period of $24$ months, or until when it has experienced a
``successful'' event (positive economic outcome) if this event occurs
before the end of the $24$ months period.
Notice however that in the greatest majority of cases firms were
observed for 24 months. For this experiment note also that we are aggregating all the firms in open-deal list in our database: not all of them are observed in the same time, e.g. one firm can be observed for the 24 months starting in January 2000, another firm can be observed for the 24 months starting in June 2004, etc. In other words, month $t=1$ for a given firm does not necessarily matches the actual date of month $t=1$ for another firm, we are simply recording the temporal evolution of different start-ups which appear in open-deal lists at different times in the period ranging from 1990 to 2008.\\ 
Then, for each of the $5,305$ firms, we conclude that a firm has  
experienced a ``successful'' event (a positive economic outcome) 
at month $t$ if, within a
time window of $\Delta t = 6, 7$ or $8$ years since month $t$, one of
the following events takes place: (i) the firm makes an acquisition;
(ii) the firm is acquired; or (iii) the firm undergoes an
IPO. Accordingly, each firm receives a unique class label (either successful with class label `1', if at any time $t+\Delta t$ the start-up experiences a successful event, or not-successful with class label '0' otherwise).
\\
Overall, this data set enables supervised learning
  (classification), as it consists of a large number of samples (the
  firms in the open-deal list), each of them described by a set of
  features (a vector of centrality measures over the whole observation
  window), and each of them being labelled by a class label.\\
   
\noindent We will use logistic regression as our supervised learning model.
   A logistic regression model links the probability of
success of a start-up to a linear combination of predictors. More precisely, 
a logistic regression model is traditionally given by:
\begin{equation}
\log \left(\frac{p_i}{1-p_i} \right) = \boldsymbol\beta^\textsc{t}\mathbf{X}_i + \varepsilon_i
\end{equation}
where $p_i$ is the probability of success of the $i$-th start-up,
$\mathbf{X}_i$ is the vector of predictors, $\boldsymbol\beta^\textsc{t}$
is the (transposed)
vector of parameters, which are estimated when the logistic regression
model is fitted, and $\varepsilon_i$ are the errors, which are assumed
to be independent, identically-distributed Normal random
variables. Rearranging terms, we have 
$$p_i = \sigma(\boldsymbol\beta^\textsc{t}\mathbf{X}_i + \varepsilon_i),$$ 
where $\boldsymbol\beta^\textsc{t}\mathbf{X}_i + \varepsilon_i$ is a linear combination of predictors with additive noise term and $\sigma(x)=1/[1+\exp(-x)]$ is the so-called logistic function.
In essence, the term $\boldsymbol\beta^\textsc{t}\mathbf{X}_i + \varepsilon_i$ is akin to a linear regression on the predictors, and the logistic function is used to force the outcome to be equal to 0 or 1:  if $p_i<c$, the class $0$ is assigned, and for $p_i>c$ the class $1$ is assigned, where the threshold $c$  is indeed another parameter that can be trained by the algorithm. Once the parameters are estimated, logistic regression can be used to predict the probability of success of new start-ups.\\
  
\noindent In what follows we consider two scenarios. In the first
  case, we define prediction models that do not consider the time evolution of centrality measures for each firm and only use instantaneous values of the firm's closeness
  centrality: these models will be closer in spirit to the recommendation method. In a second case, we
  enrich the predictor set by adding predictors summarising the time
  evolution of the firm's centrality over the observation period (to
  assess whether this factor is informative) as well as similar
  quantities extracted from different centrality measures.

\subsection{Logistic regression: the unbalanced case.}

Here we use the ROC
(receiver operating characteristic) curve to assess the efficacy of
this binary classification algorithm to choose the optimal threshold
based on our tolerance for false negatives and desire for true
positives. We initially have used only the last value of the (rescaled) closeness centrality of a start-up over the observed period as the single predictor, in order to try to match the conditions of our recommendation method where only instantaneous information is used.
The estimation and prediction
  steps above have been repeated 1,000 times, leaving out 10\% of the data set
  (Monte Carlo cross-validation). Averaging over the prediction results, we obtain
  the confusion matrix reported in Table \ref{conf-matrix} (left panel),
  together with the confusion matrix expected for a random classifier
  operating on the same data set (right panel).
\begin{table}[h]
\begin{tabular}{|c|cc|c|}
\hline
{\bf \textcolor{blue}{ACTUAL}}&&{\bf \textsf{predicted}}&\\
&&failure&success\\
\hline
{\bf \textsf{true}} &failure&0.43&0.34\\
&success&0.11&0.13\\
\hline
\end{tabular}
\begin{tabular}{|c|cc|c|}
\hline
{\bf \textcolor{blue}{RANDOM}}&&{\bf \textsf{predicted}}&\\
&&failure&success\\
\hline
{\bf \textsf{true}} &failure&0.593&0.177\\
&success&0.177&0.053\\
\hline
\end{tabular}

\caption{(Left) Confusion matrix for a logistic
    regression based on the last value of closeness and on the mean closeness
    over the entire period in the unbalanced case. Averages over
    1,000 repetitions of the Monte Carlo cross-validation leaving out 10\% of the
    data.~\label{conf-matrix} (Right) Corresponding confusion matrix
    expected for a random classifier in the same unbalanced case.}
\end{table}

Classical ways to assess the prediction performance include the
evaluation of {\it accuracy}, defined as the total percentage of
correctly identified samples, {\it sensitivity}, defined as the
percentage of successful start-ups correctly predicted by the
classifier over the percentage of true successful start-ups, and {\it
  precision}, defined as the percentage of successful start-ups
correctly predicted by the classifier over the total percentage of
start-ups which are predicted as successful by the classifier. The {\it
  F1 score} is the harmonic average of precision and
sensitivity. Depending on the context, it might be desirable for a
classifier to have high sensitivity or precision, and when both
quantities are relevant then the F1 score is typically used to assess
model selection. In our case the sensitivity is the relevant quantity
to look at if we want to maximise the detection of successful
start-ups, whereas the precision is important if we want to make sure
that all the start-ups classified as successful will be successful.
In other words, the first performance indicator can be the one of
relevance for an investment company with unlimited budget, while the
precision can be of interest to an investment company with limited
budget.

The values obtained for the different indicators are
   reported in Table \ref{summary}. The {\it F1
   score} --which trades off sensitivity and precision-- shows that the predictions on whether any start-up in
   the open-deal list will have a positive outcome are systematically better than those of a benchmark given 
   by a random classifier. Note that the problems with the accuracy are due to
   the fact that our two classes are unbalanced, and this can affect
   the usefulness of this indicator. We will come back to
   this point in the next subsection.

\noindent We have also experimented by including additional
  features of the evolution over time of the closeness centrality as predictors
  in the logistic regression model. Interestingly,
 our results did not improve significantly, suggesting that it is not necessary to use temporal evolution of centrality measures, and thus confirming the validity of the recommendation method.  This observation will be further explored in the next subsection.

\begin{table}[h]
\centering 
\begin{tabular}{|c|c|c|c|a|}
\hline
& {\bf Accuracy} & {\bf Sensitivity} & {\bf Precision} & {\bf F1 score} \\
\hline
 Unbalanced &0.56&0.54&0.28&{\bf 0.37}\\
 Unbalanced (random classifier) &0.65&0.31&0.31&{\bf 0.31}\\
\hline 
Balanced (single predictor) & 0.58 & 0.61 & 0.57 &{\bf 0.59}\\
Balanced (with temporal information) & 0.59 & 0.62 & 0.58 &{\bf 0.60}\\
Balanced (random classifier) & 0.5 & 0.5 & 0.5 &{\bf 0.5}\\
\hline
\end{tabular}
\caption{Summary of the performance indicators obtained
    for a logistic regression model to predict the success of
    start-ups in the open-deal list based on the last and on the mean value
    of closeness over time. Both unbalanced and balanced cases are considered. 
\label{summary}}
\end{table}

\subsection{Logistic regression: the balanced case}

It is well known that
many binary classification algorithms suffer if the two classes are
unbalanced, i.e. if the number of samples in each class is not
similar. A classifier would then systematically try to fit the
over-represented class and, as an outcome, the classification would be 
biased. Consider, e.g., the extreme case where the classifier assigns
each sample to the over-represented class. In this extreme situation, 
the classifier would not be predicting
anything, but the classification accuracy would still be very high
due to class unbalance. For such a reason most classifiers do not perform
well for unbalanced classes, and in unbalanced classification, accuracy can
be a misleading metric.  This is indeed our case, as in our 
data set the majority of start-ups do not end up being successful.
Here, we show that, when we correct for class unbalancing, then the
prediction performance substantially improves. In order to solve the
issue of unbalanced classes, we downsample the over-represented class,
so that the successful/non-succesful classification problem has now perfectly 
balanced ($50\%-50\%$) classes.

\noindent All over this section we use 5-fold crossvalidation. First we have considered that case where we only
  use the value of the closeness centrality of each start-up in the
  last month of our observation window, this being closer in spirit to
  the analysis performed in the main part of the manuscript.
  Again, the
  descriptor used is the closeness centrality rescaled ranking. The performance indicators of this logistic regression model 
  are reported in Table \ref{summary}, while the confusion matrix is shown in Table \ref{conf-matrix2}. Results confirm that prediction is indeed possible, and performance indicators are safely superior to random benchmarks.

 \begin{table}[h]
\begin{tabular}{|c|cc|c|}
\hline
{\bf \textcolor{blue}{ACTUAL}}&&{\bf \textsf{predicted}}&\\
&&failure&success\\
\hline
{\bf \textsf{true}} &failure&0.275&0.227\\
&success&0.193&0.305\\
\hline
\end{tabular}
\begin{tabular}{|c|cc|c|}
\hline
{\bf \textcolor{blue}{RANDOM}}&&{\bf \textsf{predicted}}&\\
&&failure&success\\
\hline
{\bf \textsf{true}} &failure&0.25&0.25\\
&success&0.25&0.25\\
\hline
\end{tabular}

\caption{(Left) Confusion matrix for a logistic
    regression with a single predictor in the balanced case , using
    5-fold cross-validation. (Right) Equivalent confusion matrix
    expected for a random classifier in the same balanced
    case.} \label{conf-matrix2}
\end{table}
 
 \noindent We have also considered
   a second logistic regression model with predictors including various statistics
   of the temporal sequence of closeness centralities in the observation
   window. We have used the following 9 predictors based on 
closeness centrality, namely:  
maximum value, minimum value, slope of a linear interpolation and 
last value of both the ranking and the rescaled ranking,
and number of months in the observation window).
The model provides an accuracy of
0.59, sensitivity 0.62 and precision 0.58, indicating that temporal
information leads to only a marginal improvement over the previous case.
\\
Finally, we have investigated other
   logistic regression models by further adding predictors related to
   other centrality measures. We find that the performance is not boosted,
   in agreement with the fact that in our case most of the other
   centrality measures tend to be correlated to the closeness,
   according to Fig.\ref{fig:SI_correlations}.
   
   \begin{table}[htb]
\centering 
\begin{tabular}{|ccc|}
\hline
3i group & advanced technology ventures & accel partners \\ 
 andreessen horowitz & atlas venture &  atomico 
 \\ august capital & austin ventures & avalon ventures \\ azure capital partners &
bain capital ventures &
balderton capital \\
battery ventures &
benchmark &
bessemer venture partners \\
binary venture partners &
canvas venture fund &
carmel ventures \\
charles river ventures &
clearstone venture partners & columbus nova \\
costanoa venture capital & crosslink capital &
crunchfund \\
data collective&
digital sky technologies fo&
draper fisher jurvetson\\
elevate ventures&
ff venture capital&
fidelity ventures\\
firstmark capital&
first round capital&
flybridge capital\\
foundation capital&
founders fund&
general catalyst partners\\
genesis partners&
golden gate capital&
ggv capital\\
google ventures&
granite ventures&
greylock partners israel\\
harris harris group&
highland capital partners&
idg ventures europe\\
idg ventures india&
idg ventures vietnam&
initial capital 2\\
in q tel&
index ventures&
innovacom\\
insight venture partners&
intel capital&
intellectual ventures\\
institutional venture partners&
inventus capital partners&
jerusalem venture partners\\
jmi equity&
kapor capital&
kleiner perkins caufield byers\\
khosla ventures&
lightspeed venture partners&
lux capital\\
matrix partners&
maveron&
mayfield fund\\
menlo ventures&
meritech capital partners&
merus capital\\
morgenthaler ventures&
new enterprise associates&
norwest venture partners\\
oak investment partners&
oregon angel fund&
openview venture partners\\
polaris partners&
radius ventures&
redpoint ventures\\
revolution capital partners&
rho ventures&
finisterre capital\\
rre ventures&
rothenberg ventures&
sante ventures\\
scale venture partners&
scottish investment bank&
scottish equity partners\\
sequoia capital&
seventure partners&
sevin rosen funds\\
the social capital partnership&
sofinnova partners&
spark capital\\
tenaya capital&
third rock ventures&
tribeca global investments\\
union square ventures&
us venture partners&
vantagepoint capital partners\\
venrock&
wellington partners&\\
\hline
\end{tabular}
\caption{List of $101$ venture capital funds extracted from \texttt{crunchbase.com}. Also available at \url{https://en.wikipedia.org/wiki/List_of_venture_capital_firms}.}  \label{table:VC}
\end{table}

\end{document}